\documentclass[usenatbib,letterpaper]{mn2e}
\usepackage[totalwidth=480pt,totalheight=680pt,twoside=false,voffset=.5cm]{geometry}
\usepackage{graphics,graphicx,epsfig,ulem,multirow,appendix,lscape,amsmath,amssymb,color,enumitem,natbib,url}

\usepackage{upgreek}
\usepackage[none]{hyphenat}	

\long\def\symbolfootnote[#1]#2{\begingroup%
\def\thefootnote{\fnsymbol{footnote}}\footnote[#1]{#2}\endgroup} 

\begin{document}

\title[Resonant debris and Fomalhaut b]
  {Fomalhaut b could be massive and sculpting the narrow, eccentric debris disc, if in mean-motion resonance with it}
\author[T. D. Pearce et al.]
  {Tim D. Pearce${^1}$\thanks{timothy.pearce@uni-jena.de},
  Herv\'{e} Beust${^2}$,
  Virginie Faramaz${^3}$,
  Mark Booth${^1}$,\newauthor
  Alexander V. Krivov${^1}$,
  Torsten L\"{o}hne${^1}$,
  Pedro P. Poblete${^1}$\\
  ${^1}$Astrophysikalisches Institut und Universit\"{a}tssternwarte, Friedrich-Schiller-Universit\"{a}t Jena, Schillerg\"{a}{\ss}chen 2-3, D-07745 Jena, Germany\\
    $^2$Univ. Grenoble Alpes, CNRS, IPAG, 38000 Grenoble, France\\
    $^3$Jet Propulsion Laboratory, California Institute of Technology, 4800 Oak Grove drive, Pasadena CA 91109, USA}
\date{Released 2002 Xxxxx XX}
\pagerange{\pageref{firstpage}--\pageref{lastpage}} \pubyear{2002}

\def\LaTeX{L\kern-.36em\raise.3ex\hbox{a}\kern-.15em
    T\kern-.1667em\lower.7ex\hbox{E}\kern-.125emX}

\newtheorem{theorem}{Theorem}[section]

\label{firstpage}

\maketitle


\begin{abstract}   

\noindent The star Fomalhaut hosts a narrow, eccentric debris disc, plus a highly eccentric companion Fomalhaut b. It is often argued that Fomalhaut b cannot have significant mass, otherwise it would quickly perturb the disc. We show that material in internal mean-motion resonances with a massive, coplanar Fomalhaut b would actually be long-term stable, and occupy orbits similar to the observed debris. Furthermore, millimetre dust released in collisions between resonant bodies could reproduce the width, shape and orientation of the observed disc. We first re-examine the possible orbits of Fomalhaut b, assuming that it moves under gravity alone. If Fomalhaut b orbits close to the disc midplane then its orbit crosses the disc, and the two are apsidally aligned. This alignment may hint at an ongoing dynamical interaction. Using the observationally allowed orbits, we then model the interaction between a massive Fomalhaut b and debris. Whilst most debris is unstable in such an extreme configuration, we identify several resonant populations that remain stable for the stellar lifetime, despite crossing the orbit of Fomalhaut b. This debris occupies low-eccentricity orbits similar to the observed debris ring. These resonant bodies would have a clumpy distribution, but dust released in collisions between them would form a narrow, relatively smooth ring similar to observations. We show that if Fomalhaut b has a mass between those of Earth and Jupiter then, far from removing the observed debris, it could actually be sculpting it through resonant interactions. 

\end{abstract}



\section{Introduction}
\label{sec: Introduction}

\noindent When a debris disc is found to be eccentric, the most commonly cited explanation is sculpting by a planet (e.g. \citealt{Wyatt1999}). The spectacular dust ring around the star Fomalhaut is no exception, and its morphology was soon ascribed to perturbations by unseen bodies \citep{Stapelfeldt2004, Marsh2005, Kalas2005, Quillen2006}. When Fomalhaut b was discovered just interior to the belt in projection, it seemed that the picture was almost complete; Fomalhaut b appeared to be the predicted planet, moving on a shepherding orbit near the disc inner edge \citep{Kalas2008, Chiang2009}. However, this neat explanation quickly faced two challenges, and the hypothesis that a low-eccentricity, planetary-mass Fomalhaut b sculpts the disc started to look doubtful.

The first issue is that Fomalhaut b does not look like a typical planet; specifically, it is much dimmer in thermal emission than would be expected from its brightness in visible light \citep{Kalas2008, Marengo2009, Janson2012}. This has led to intense discussions about what Fomalhaut b actually is. Suggestions include a planet with a ring system \citep{Kalas2008}, or surrounded by a swarm of collisional satellites \citep{Kennedy2011, Currie2012, Galicher2013, Kenyon2014, Tamayo2014}. It may not be planetary at all, but rather a transient, dispersing dust cloud \citep{Janson2012, Galicher2013, Lawler2015, Gaspar2020, Janson2020}, or even a background object not associated with the system (\citealt{Neuhauser2015}; see also \citealt{Poppenhaeger2017}). Given this diversity of possibilities, the mass of Fomalhaut b, and hence its ability to sculpt the disc, is highly uncertain.

The second issue is that the sky-plane motion of Fomalhaut b is inconsistent with a low-eccentricity orbit \citep{Kalas2013, Beust2014, Pearce2015OrbCnstrnts, Gaspar2020}. Furthermore, if Fomalhaut b is coplanar with the disc midplane and moves under gravity alone, then it must pass \textit{through} the disc. A highly eccentric Fomalhaut b cutting across the disc is a more extreme orbital configuration than the low-eccentricity, shepherding planet originally proposed, and any ensuing dynamical interaction would be more complex. Indeed, various works modelled similar setups, and found that long-term debris evolution driven by secular and scattering interactions results in a debris morphology very different to the observed disc \citep{Beust2014, Tamayo2014}. This gives rise to the commonly argued suggestion that Fomalhaut b cannot be massive enough to significantly perturb debris, because if it were, then it would rapidly disrupt the observed dust ring.

However, if Fomalhaut b moves in or close to the disc midplane, then its orbit is near apsidal alignment with the disc \citep{Kalas2013, Beust2014, Pearce2015OrbCnstrnts}. Since coplanarity seems likely from system formation and evolution arguments, it appears probable that the orbit of Fomalhaut b (if bound) is indeed well-aligned with the disc. Such alignment is typically expected from long-term dynamical interactions (e.g. \citealt{Murray1999}). If Fomalhaut b really cannot be massive enough to perturb the disc, then this apparent alignment is either coincidence, evidence for the disc perturbing Fomalhaut b, or evidence for another body perturbing both the disc and Fomalhaut b \citep{Faramaz2015}. Alternatively, perhaps Fomalhaut b \textit{is} massive, and is sculpting the disc through some unexplored dynamical interaction.

In this paper we challenge the argument that Fomalhaut b cannot have significant mass without disrupting the observed disc, by considering debris in \textit{internal mean-motion resonances} with Fomalhaut b. Resonances in the Fomalhaut system have been discussed before \citep{Wyatt2002, Holland2003, Chiang2009, Beust2014}, but only in the context of either low-eccentricity perturbers or non-internal resonances. We consider observationally allowed orbits of Fomalhaut b, and show that resonant debris on orbits similar to the observed disc is stable for the stellar lifetime, despite crossing the orbit of a massive, highly eccentric perturber. If this scenario were operating in the Fomalhaut system, then these resonant objects would be the large parent bodies in the disc. Debris with similar but non-resonant configurations is unstable. We also show that dust produced by collisions between these resonant bodies could have a similar shape, width and orientation to the Fomalhaut disc, even though these grains are not necessarily resonant themselves. Note that the aim of this paper is not to determine the nature of Fomalhaut b, nor to constrain its orbit or history, but rather to show that a massive Fomalhaut b on an observationally allowed, disc-crossing orbit does not necessarily disrupt the observed ring. This potentially has implications beyond the Fomalhaut system; an increasing number of narrow, eccentric debris discs are now known (e.g. ${\rm HR \; 4796A}$, \citealt{Kennedy2018}; ${\rm HD \; 202628}$, \citealt{Faramaz2019}), and the scenario we present provides a new possible means for a planet to sculpt them.

The structure of the paper is as follows. In Section \ref{sec: possibleOrbits} we examine all possible bound orbits of Fomalhaut b assuming it moves under gravity alone, including newly reduced data from \cite{Gaspar2020}. In Section \ref{sec: planetDiscInteraction} we investigate the interaction between debris and a massive Fomalhaut b on some of the orbits from Section \ref{sec: possibleOrbits}, and show that resonant bodies on similar orbits to the observed disc can be stable for the stellar lifetime. In Section \ref{sec: fomalhautDiscAsResRing} we show that, provided there is at least one additional planet in the system, the millimetre dust produced by collisions between resonant bodies could provide a good match to the observed debris disc. We discuss our results in Section \ref{sec: discussion}, and conclude in Section \ref{sec: conclusions}.


\section{All possible bound orbits of a massive Fomalhaut \lowercase{b}}
\label{sec: possibleOrbits}

\begin{figure}
  \centering
   \includegraphics[width=5cm]{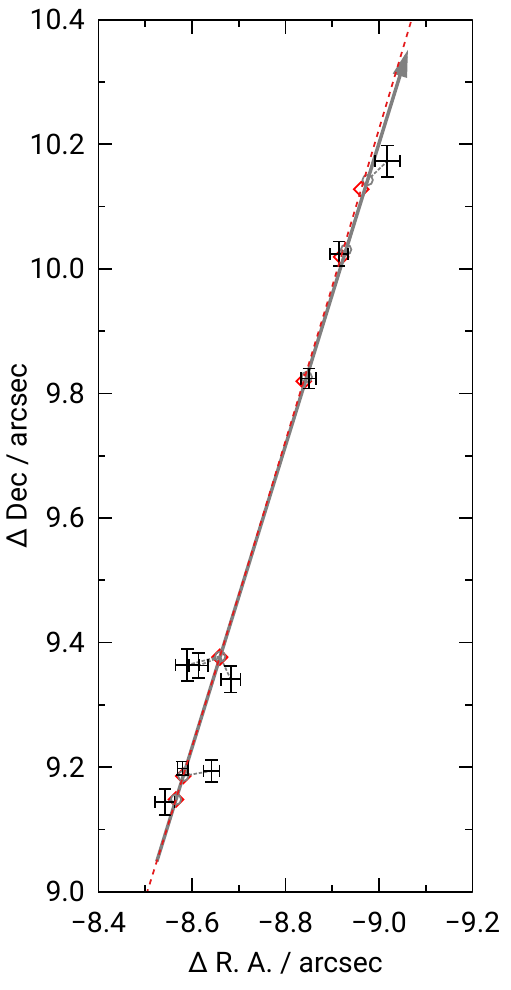}
   \caption{Observed positions of Fomalhaut b since 2004, relative to the star. North is up, east is left. Data points are listed in Table \ref{tab: fomBObservedPositions}. The solid grey line shows the expected trajectory if Fomalhaut b moves in a straight line at a constant velocity, in the direction of the arrow, and circles show its expected position at each observation epoch. The red dashed line and diamonds show the predicted positions for one particular orbit with semimajor axis ${203 \; \rm au}$ and eccentricity 0.760, as used in the dynamical simulations shown on Figures \ref{fig: fomalhautbElementsInRegionOfInterest}-\ref{fig: collisionsHeatmap}; note that many other orbits also satisfy these observational constraints.}
   \label{fig: fombMotionFit}
\end{figure}

Before exploring any dynamical interactions, we must first establish what orbits of Fomalhaut b are observationally allowed. The confirmed positions of Fomalhaut b relative to the star are listed in Table \ref{tab: fomBObservedPositions}, and shown on Figure \ref{fig: fombMotionFit}. No orbital curvature is apparent. We therefore fit the observed sky positions with a model for an object moving in a straight line at constant velocity, and find that the observations are well-approximated by a body moving north at $0.11 \; {\rm arcsec \; yr}^{-1}$ and west at $0.046 \; {\rm arcsec \; yr}^{-1}$ relative to the star. This fit is shown on Figure \ref{fig: fombMotionFit}.

\begin{table*}
\begin{tabular}{c c c c c}
\hline
Date & Instrument & $\Delta$ R. A. / arcsec & $\Delta$ Dec. / arcsec & Reference\\
\hline
25/06/2004	 & ACS (F814W) & $-8.542 \pm 0.021$ & $9.144 \pm 0.021$ & 1\\
27/10/2004	 & ACS (F606W) & $-8.580 \pm 0.011$ & $9.198 \pm 0.011$ & 1\\
27/10/2004 & ACS (F814W) & $-8.642 \pm 0.017$ & $9.194 \pm 0.018$ & 1\\
15/07/2006	 & ACS (F435W) & $-8.614 \pm 0.020$ & $9.363 \pm 0.020$ & 2\\
17/07/2006	 & ACS (F606W) & $-8.683 \pm 0.021$ & $9.341 \pm 0.021$ & 2\\
19/07/2006	 & ACS (F814W) & $-8.590 \pm 0.025$ & $9.364 \pm 0.026$ & 2\\
29/07/2010	 & STIS	 & $-8.850 \pm 0.016$ & $9.824 \pm 0.016$ & 3\\
30/05/2012	 & STIS	 & $-8.915 \pm 0.019$ & $10.024 \pm 0.020$ & 3\\
31/05/2013	 & STIS	 & $-9.018 \pm 0.027$ & $10.173 \pm 0.025$ & 4\\
\hline
\end{tabular}
\caption{Observations of Fomalhaut b where the object was detected. Dates are the approximate midpoint of each observation. Positions are relative to the star, from the data re-reduction by \citet{Gaspar2020}. References are the publications that first presented the observations: (1) \citet{Kalas2005}, (2) \citet{Kalas2008}, (3) \citet{Kalas2013}, (4) \citet{Gaspar2020}.}
\label{tab: fomBObservedPositions}
\end{table*}

Since sky-plane positions without curvature cannot uniquely determine the orbit of an imaged companion, the observations shown on Figure \ref{fig: fombMotionFit} are consistent with a wide range of bound and unbound orbits. We now examine all possible bound orbits of Fomalhaut b, assuming it moves under gravity alone. Whilst it has been suggested that Fomalhaut b shows non-gravitational acceleration \citep{Gaspar2020}, the uncertainties on its observed positions and the small size of the proposed deviation make this uncertain without more data. Since we investigate a scenario involving a planetary-mass Fomalhaut b, for which non-gravitational forces would be negligible, we model its orbit under gravity alone; we discuss this assumption in relation to the nature of Fomalhaut b in Section \ref{subsec: natureOfFomb}.

In order to proceed we require the star parameters, plus a reference plane and direction from which to define an orbit orientation. For the latter we use the plane and pericentre direction of an elliptical orbit that traces the centre of the ALMA-resolved debris disc \citep{Macgregor2017}. This orbit is defined by the parameters ${a_{\rm disc}}$, ${e_{\rm disc}}$, ${i_{\rm disc}}$, ${\Omega_{\rm disc}}$ and ${\omega_{\rm disc}}$, which denote the semimajor axis, eccentricity, inclination (relative to the sky plane), longitude of ascending node (measured from north and turning positive towards east, where a body crossing the sky plane would move towards Earth) and argument of pericentre (measured from the ascending node) of the centre of the disc, respectively. However, since the orbit would appear identical from Earth if ${\Omega_{\rm disc}}$ and ${\omega_{\rm disc}}$ were replaced by ${\Omega_{\rm disc} + 180^\circ}$ and ${\omega_{\rm disc} + 180^\circ}$ respectively \citep{Beust2014, Pearce2015OrbCnstrnts}, we also need to decide which side of the disc lies in front of the star. We assume the simplest dynamical argument, that if Fomalhaut b lies close to the disc plane then the star, disc and Fomalhaut b all rotate/orbit in the same direction. Coupling the sky-plane motion of Fomalhaut b with the stellar rotation axis \citep{LeBouquin2009} and the motion of gas in the disc \citep{Matra2017}, this would make the north-west side of the disc closer to Earth than the south-east side. A coplanar Fomalhaut b would therefore lie on the near side of the star as viewed from Earth. Note that this disc orientation is the opposite to that assumed in \citet{Kalas2013}, and would have implications for the size of dust grains in the ring \citep{Min2010}. However, making the opposite orientation assumption would not affect any orbital elements of a coplanar Fomalhaut b, other than its longitude of ascending node and argument of pericentre each changing by ${180^\circ}$. The star mass ${M_*}$, age ${t_*}$ and distance ${d}$ are listed in Table \ref{tab: starAndDiscParameters}, along with the parameters of the orbit tracing the disc centre according to the above assumptions. Also listed are ${\Delta a_{\rm disc}}$, the range of debris semimajor axes, and ${\Delta r_{\rm disc}}$, the Full Width Half Maximum (FWHM) disc width.

\begin{table}
\begin{tabular}{l l c}
\hline
Parameter & Value & Reference \\
\hline

${M_*}$ & ${1.92 \pm 0.02 \; {\rm M_\odot}}$ & 1\\
${t_*}$ & ${440 \pm 40 \; {\rm Myr}}$ & 1\\
${d}$ & ${7.70 \pm 0.03 \; {\rm pc}}$ & 2\\
${a_{\rm disc}}$ & ${^{a}142.4 \pm 1.2 \; {\rm au}}$ & 3\\
${e_{\rm disc}}$ & ${0.12 \pm 0.01}$ & 3\\
${i_{\rm disc}}$ & ${^{b}65.6 \pm 0.3^\circ}$ & 3\\
${\Omega_{\rm disc}}$ & ${^{c}157.9 \pm 0.3^\circ}$ & 3\\
${\omega_{\rm disc}}$ & ${^{d}22.5 \pm  4.3^\circ}$ & 3\\
${\Delta a_{\rm disc}}$ & ${12.2 \pm 1.6 \; {\rm au}}$ & 3\\
${\Delta r_{\rm disc}}$ & ${13.5 \pm 1.8 \; {\rm au}}$ & 3\\
\hline
\end{tabular}\caption{Parameters of Fomalhaut A and disc used in this paper. The parameters ${a_{\rm disc}}$, ${e_{\rm disc}}$, ${i_{\rm disc}}$, ${\Omega_{\rm disc}}$ and ${\omega_{\rm disc}}$ describe an elliptical orbit tracing the centre of the resolved debris disc. ${\Delta a_{\rm disc}}$ is the range of disc semimajor axes, and ${\Delta r_{\rm disc}}$ is the FWHM disc width. Notes: ${^{a}}$Derived from parameters in \citet{Macgregor2017}; ${^{b}}$relative to the sky plane; ${^{c}}$measured from north and turning positive towards east, assuming the north-west side of the disc is closer to Earth than the south-east side; ${^{d}}$measured from the ascending node. References: (1) \citealt{Mamajek2012}, (2) \citealt{vanLeeuwen2007}, (3) \citealt{Macgregor2017}.}
\label{tab: starAndDiscParameters}
\end{table}

\subsection{All possible orbits}
\label{subsec: allPossibleOrbits}

\begin{figure*}
  \centering
   \includegraphics[width=16cm]{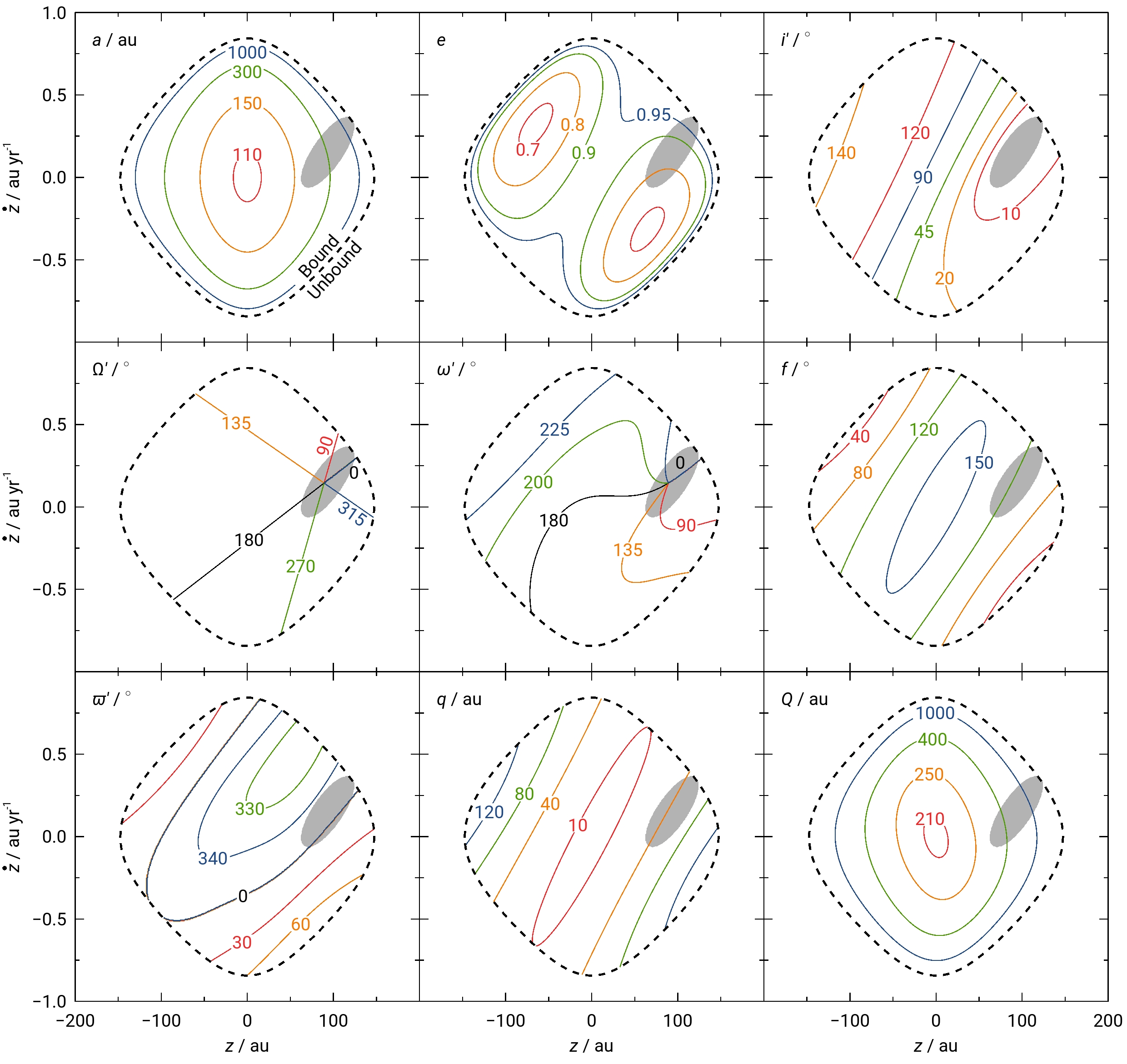}
   \caption{All possible bound orbits of Fomalhaut b, as functions of its unknown line-of-sight position $z$ and velocity $\dot{z}$. Each ${(z, \dot{z})}$ pair yields a unique orbital solution, with orbital elements that can be read off the plots. Orbits inside the dashed black lines are bound to the star, whilst those outside are unbound. The shaded grey regions are prograde orbits within $5^\circ$ of the disc midplane, and primed parameters denote those relative to the disc. If Fomalhaut b orbits within $5^\circ$ of the disc midplane, then it is likely bound to the star with eccentricity ${e \geq 0.76}$. It would also be apsidally aligned with the disc, with a maximum orbital misalignment of ${\left|\varpi'\right| \leq 17^\circ}$. Earth lies along positive $z$.}
   \label{fig: fomalhautbAllowedOrbits}
\end{figure*}

\noindent Whilst the sky-plane position and velocity of Fomalhaut b are well-constrained, no information is known about its line of sight position $z$ and velocity $\dot{z}$. Since 6 coordinates are required to uniquely constrain an orbit (4 from the sky-plane position and velocity, plus $z$ and $\dot{z}$), every pair of possible line-of-sight coordinates yields a unique orbital solution. Using the method of \citet{Pearce2015OrbCnstrnts}, by considering a wide range of line-of-sight positions and velocities we can therefore explore the possible orbits of Fomalhaut b, assuming that it moves under gravity alone. We generate a regular grid of 400 $z$ values and 400 $\dot{z}$ values, where the ranges ${|z| \leq 200 \; \rm au}$ and ${|\dot{z}| \leq 1 \; \rm au \; yr^{-1}}$ are considered\footnote{Given the sky-plane motion of Fomalhaut b, if it moves under gravity alone then it can only be bound if ${|z| \leq 150 \; \rm au}$ and ${|\dot{z}| \leq 0.85 \; \rm au \; yr^{-1}}$ (equation 4 in \citealt{Pearce2015OrbCnstrnts}).}. For each ${(z, \dot{z})}$ pair we calculate the orbital solution using the system parameters in Table \ref{tab: starAndDiscParameters}, combined with the fitted sky-plane position and velocity at the time of the first observation. We show all possible bound orbits on Figure \ref{fig: fomalhautbAllowedOrbits}, where as a function of the unknown line-of-sight coordinates we plot Fomalhaut b's semimajor axis $a$, eccentricity $e$, inclination $i'$ (relative to the disc plane), longitude of ascending node $\Omega'$ (measured from the disc pericentre), argument of pericentre $\omega'$, true anomaly $f$, longitude of pericentre $\varpi' \equiv \Omega' + \omega'$, pericentre distance $q$ and apocentre distance $Q$; primes denote values relative to the disc. Note that the presented true anomaly is that at the first observation epoch (${25/06/2004}$).

The possible orbits of Fomalhaut b have been explored before, and the addition of the new 2013 detection by \citet{Gaspar2020} does not significantly affect the conclusions. If bound, Fomalhaut b must have a semimajor axis of at least ${107 \; \rm au}$, an eccentricity of at least ${0.68}$, and currently be moving from pericentre to apocentre (${29^\circ \leq f \leq 160^\circ}$), in line with previous studies \citep{Kalas2013, Beust2014, Pearce2015OrbCnstrnts}. Its apocentre must be at least ${200 \; {\rm au}}$, which is exterior to the disc, and its pericentre may be as small as ${7 \; {\rm au}}$. Its orbital plane and orientation are relatively unconstrained. Note that it is always possible for Fomalhaut b to be unbound, since there are no limits on its unknown line-of-sight position or velocity.

\subsection{Orbits close to the disc plane}
\label{subsec: orbitsCloseToDiscPlane}

\noindent Having examined all possible bound orbits of Fomalhaut b, we now focus on those that are close to the disc plane and prograde relative to the disc. Such orbits could be expected from system formation and evolution arguments. All orbits within $5^\circ$ of the disc plane are shown as the grey regions on Figure \ref{fig: fomalhautbAllowedOrbits}; these orbits are much less diverse than those allowed if Fomalhaut b is not assumed to lie close to the disc plane, with a narrower range of possible orbital elements.

If Fomalhaut b orbits within $5^\circ$ of the disc plane and moves under gravity alone, then it is very likely bound to the star (although very high eccentricities are possible). This is apparent from Figure \ref{fig: fomalhautbAllowedOrbits}, where the grey regions (showing orbits within $5^\circ$ of the disc plane) do not extend beyond the black dashed lines (separating bound and unbound orbits). Its semimajor axis would be at least ${170 \; \rm au}$, its eccentricity at least $0.76$, and its true anomaly constrained to ${100^\circ \leq f \leq 130^\circ}$. Perhaps most interestingly, its pericentre would be very well-aligned with that of the disc (${-17^\circ \leq \varpi' \leq 3^\circ}$), an outcome often associated with long-term dynamical interactions (e.g. \citealt{Murray1999}). Fomalhaut b would currently be on the near side of the star as viewed from Earth (${65 \; {\rm au} \leq z \leq 120 \; {\rm au}}$), and most orbital solutions have it moving towards us (${-0.05 \; {\rm au \; yr^{-1}} \leq \dot{z} \leq 0.35 \; {\rm au \; yr^{-1}}}$).

\begin{figure}
  \centering
   \includegraphics[width=7cm]{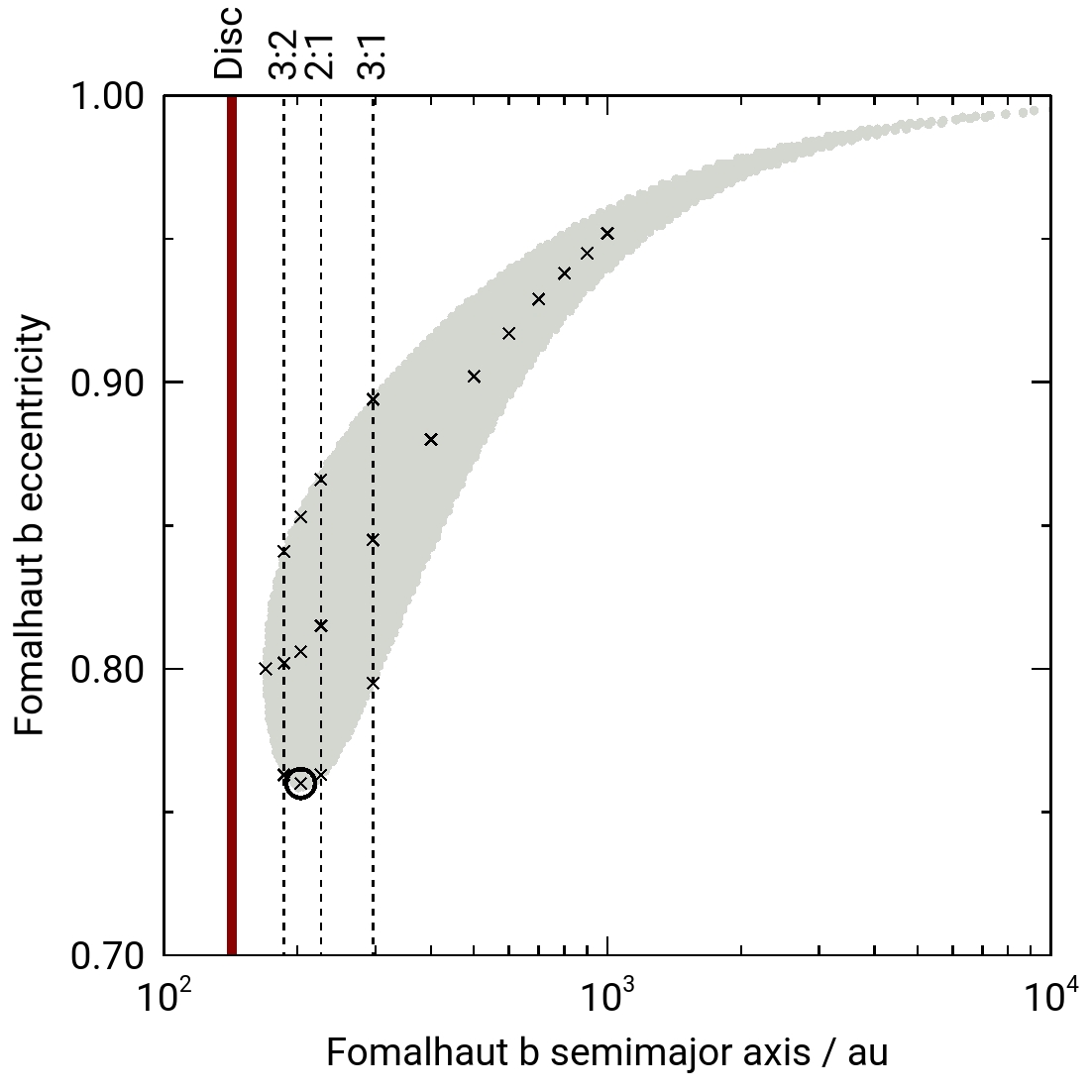}
   \caption{The grey region shows all possible semimajor axes and eccentricities of Fomalhaut b, if it orbits within ${5^\circ}$ of the disc midplane and moves under gravity alone. Each grey dot is an orbital solution for a tested pair of line-of-sight coordinates; increasing the resolution of the tested parameter space would fill in any spaces between dots, but leave the overall shape of the plot unchanged. The solid brown line shows the central semimajor axis of the observed disc, and dashed lines show several possible semimajor axes of Fomalhaut b that would result in the disc being at the nominal location of an internal mean-motion resonance (for example `$3{:}2$' means that the disc would complete 3 orbits for every 2 orbits of Fomalhaut b). If Fomalhaut b orbits within ${5^\circ}$ of the disc midplane then its semimajor axis must be greater than that of the disc, in which case the 1:1 and external resonances are not possible. Crosses denote orbits of Fomalhaut b that we examine in dynamical simulations (Section \ref{sec: planetDiscInteraction}), and the black circle marks the example orbit with semimajor axis ${203 \; \rm au}$ and eccentricity 0.760 that we show on Figures \ref{fig: fombMotionFit} and \ref{fig: generalSimPosAE}-\ref{fig: collisionsHeatmap}.}
   \label{fig: fomalhautbElementsInRegionOfInterest}
\end{figure}

Fomalhaut b must also cross the disc if the two orbit within $5^\circ$ of each other. In this case the apocentre of Fomalhaut b would be at least ${300 \; \rm au}$, well-exterior to the disc semimajor axis of ${142 \; \rm au}$, whilst its pericentre would be interior to the disc with a maximum value of ${65 \; \rm au}$. Its smallest possible pericentre would be ${30 \; \rm au}$. Its semimajor axis would also be larger than that of the disc, which is important for later discussions of mean-motion resonances; on Figure \ref{fig: fomalhautbElementsInRegionOfInterest} we plot the possible semimajor axes and eccentricities of Fomalhaut b if it orbits close to the disc plane, and show the nominal locations of several internal resonances.

\section{Planet - disc interaction}
\label{sec: planetDiscInteraction}

\noindent In Section \ref{subsec: orbitsCloseToDiscPlane} we showed that, if Fomalhaut b is coplanar with the disc and moves under gravity alone, then it must be on an eccentric, disc-crossing orbit which is apsidally aligned with the disc. Such apsidal alignment could be pointing to some dynamical interaction. In this section we explore a scenario where Fomalhaut b is massive and coplanar, and sculpts the observed disc through an ongoing interaction that ensures the long-term stability of debris.

An interaction would be scattering, secular or resonant in nature, or some combination of these. Scattering alone can be discounted because it would not form a narrow, coherently eccentric debris ring, but rather a cloud of scattered objects with  a very broad range of eccentricities and orientations. Likewise, whilst secular interactions can produce eccentric debris rings that are long-term stable and apsidally aligned with the perturber (e.g. \citealt{Faramaz2014, Pearce2014}), this behaviour is not predicted for high-eccentricity perturbers on debris-crossing orbits; such configurations initially lead to misaligned debris rings, which then evolve into a `V'-shaped distribution rather than a single ring (as will be described in Section \ref{subsec: numericalResultsAllDebris}; \citealt{Beust2014, Tamayo2014}). Therefore, since scattering and secular interactions between a coplanar Fomalhaut b and the disc are unlikely to result in the observed debris ring, we focus on mean-motion resonances for the rest of this investigation.

We describe the scenario considered in Section \ref{subsec: scenario}, and explore this numerically in Section \ref{subsec: numericalInvestigationAllDebris}. In Section \ref{subsec: numericalInvestigationStableOnly} we perform a more detailed, specific investigation into long-term-stable debris, and show that resonant debris can occupy orbits similar to the observed disc, despite crossing the orbit of a coplanar, massive Fomalhaut b.

\subsection{Scenario}
\label{subsec: scenario}

\noindent We explore a scenario where an eccentric Fomalhaut b interacts with an initially coplanar, unexcited, axisymmetric debris disc, and compare any surviving debris to the observed ring. The model initially consists of the star, a massive Fomalhaut b, and debris, with no other bodies included. This debris represents collisionless parent bodies; we will investigate collisions and the subsequent dust release in Section \ref{subsec: gapsAndCollisions}. We consider only observationally allowed orbits of Fomalhaut b that lie within $5^\circ$ of the disc plane (Section \ref{subsec: orbitsCloseToDiscPlane}), and mainly investigate orbits that place the observed disc close to a major internal mean-motion resonance.

If Fomalhaut b has semimajor axis $a_{\rm plt}$, then the nominal location of the ${p+q:p}$ internal resonance is

\begin{equation}
a_{\rm res} \approx a_{\rm plt} \left(\frac{p+q}{p} \right)^{-2/3};
\label{eq: nominalResonanceLocation}
\end{equation}

\noindent a particle with semimajor axis $a_{\rm res}$ would complete roughly $p+q$ integer orbits for every $p$ integer orbits completed by the putative planet ($q$ is referred to as the order of the resonance). We mainly consider semimajor axes for Fomalhaut b that result in ${a_{\rm res} \approx 142.4 \; {\rm au}}$, the central semimajor axis of the observed ring (Table \ref{tab: starAndDiscParameters}). Since the scenario starts with a coplanar axisymmetric disc, the only other orbital element of Fomalhaut b that determines the long-term interaction is its eccentricity; for each semimajor axis considered we pick an eccentricity from the allowed orbits that lie within $5^\circ$ of the disc plane. Finally, since the mass of Fomalhaut b is poorly constrained from observations, we consider planet masses of ${M_{\rm plt} = 1}$, 0.1 and $0.01 \; {\rm M_{Jup}}$ (300, 30 and ${3 \; {\rm M_\oplus}}$, respectively).

\subsection{Numerical investigation of all debris}
\label{subsec: numericalInvestigationAllDebris}

\noindent We first use n-body simulations to investigate the scenario described above. We describe the general simulation setup in Section \ref{subsec: numericalSetupAllDebris}, and the results in Section \ref{subsec: numericalResultsAllDebris}.

\subsubsection{Setup of general n-body simulations}
\label{subsec: numericalSetupAllDebris}

\noindent For each simulation in this section we include the star, a massive Fomalhaut b and ${\sim 1000}$ debris particles. The semimajor axes and eccentricities of Fomalhaut b in our various simulations are marked by crosses on Figure \ref{fig: fomalhautbElementsInRegionOfInterest}; most of these orbits would place the disc near the strong 3:2, 2:1 or 3:1 internal resonances, and we probe a range of eccentricities for Fomalhaut b at those semimajor axes. We also test other semimajor axes for Fomalhaut b, up to ${1000 \; \rm au}$. The parameters of Fomalhaut b in each simulation are listed in Table \ref{tab: simulationsRun} in Appendix \ref{app: simulationsRun}.

For each simulation the initial disc is set up in a similar manner to \citet{Pearce2014, Pearce2015PltDsc}. Each debris particle has an initial semimajor axis $a$ drawn from the distribution

\begin{equation}
n(a) \propto a^{1-\gamma}
\label{eq: initialDiscSemimajorAxisDistribution}
\end{equation}

\noindent with ${\gamma = 1.5}$, which is the surface density index of the Minimum Mass Solar Nebula \citep{Weidenschilling1977, Hayashi1981}. The initial semimajor axis range spanned by our discs is ${125-159 \; \rm au}$, which is substantially broader than the observed debris ring around Fomalhaut. It also means that all debris in our simulations is initially located between pericentre and apocentre of Fomalhaut b. The initial eccentricity $e$ of each particle is uniformly drawn in the range ${0 \leq e \leq 0.05}$; for a resonant particle the initial eccentricity is important, because it can set the behaviour of the body in ${e-\varpi}$ space as discussed in Section \ref{subsec: evolutionOfResonantParticles}. The inclination $i$ of each particle is uniformly drawn in the range ${0^\circ \leq i \leq 5^\circ}$, and its longitude of ascending node $\Omega$, argument of pericentre $\omega$ and mean anomaly are each uniformly drawn between 0 and $360^\circ$. Debris particles are massless in our simulations, so the disc does not experience self-gravity or perturb the orbit of Fomalhaut b; the consequences of this assumption are discussed in Section \ref{subsec: discMass}. We also omit radiation forces, since the simulated debris represents large parent bodies (we will discuss smaller dust grains in Section \ref{subsec: gapsAndCollisions}).

Simulations are performed in {\sc rebound} with the {\sc ias15} integrator \citep{Rein2012, ReinIAS152015}, and the main results checked with {\sc mercury6} using the {\sc hybrid} integrator \citep{Chambers1999}. Simulations are run for ${440 \; {\rm Myr}}$, the lifetime of the system, and particles are only removed if their stellocentric distance exceeds ${10^4 \; {\rm au}}$.

\subsubsection{Results of general n-body simulations}
\label{subsec: numericalResultsAllDebris}

\noindent The results of the general simulations are qualitatively similar across the whole parameter space, and we describe the common outcomes here. The majority of debris initially present in the scenario is unstable over the lifetime of the star; this is unsurprising given the extreme initial configuration, in which all particles cross the orbit of a highly eccentric perturber. Despite this, some debris does occupy low-eccentricity orbits that are stable over the system lifetime, and we will examine these populations in detail in Section \ref{subsec: numericalInvestigationStableOnly}. In all of our simulations, secular, scattering and resonant effects all manifest themselves.

\begin{figure*}
  \centering
   \includegraphics[width=16cm]{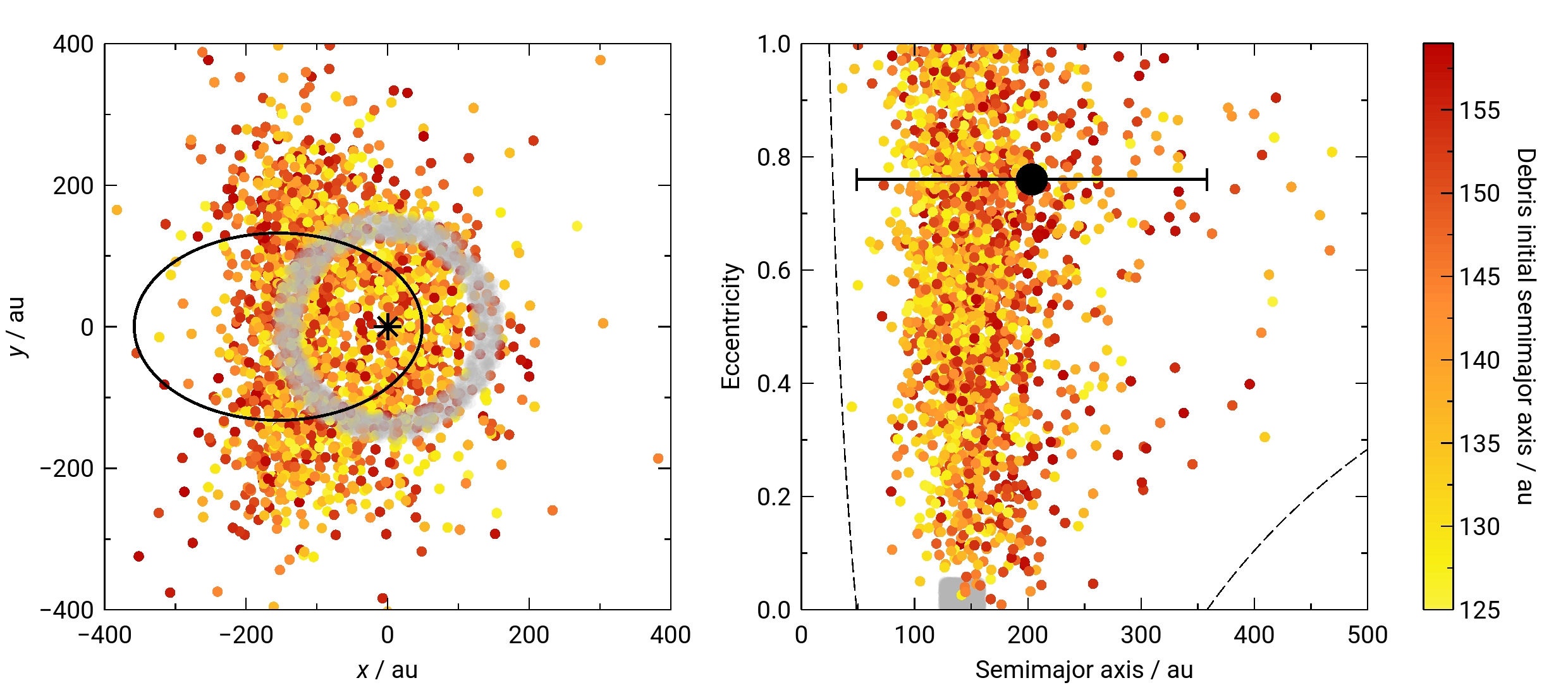}
   \caption{Parameters of debris during an interaction with a ${0.1 \; {\rm M_{Jup}}}$ (${30 \; {\rm M_\oplus}}$) Fomalhaut b, shown after ${440 \; \rm Myr}$ (the age of the system). The simulation is described in Section \ref{subsec: numericalInvestigationAllDebris}; here Fomalhaut b has a semimajor axis of ${203 \: {\rm au}}$ and an eccentricity of $0.760$. Debris particles are the red-yellow points, coloured by initial semimajor axis. They started on low-eccentricity, low-inclination orbits with semimajor axes between 125 and 159 au, as shown by the grey shaded regions. The asterisk and black ellipse on the left plot denote the star and orbit of Fomalhaut b respectively. On the right plot the black point is Fomalhaut b, and the solid line shows its pericentre-apocentre span. Particles above the dashed lines on the right plot can be on orbits which intersect that of Fomalhaut b. Secular interactions cause the eccentricities of most debris particles to oscillate from ${\sim 0}$ to ${\sim 1}$ whilst conserving semimajor axis, driving the V-shaped structure on the left plot. Close encounters with Fomalhaut b cause particles to diffuse in semimajor axis space. The majority of particles are unstable, but there are also some long-term-stable, low-eccentricity, resonant populations hidden in the plot, as described in Section \ref{subsec: numericalInvestigationStableOnly}.}
   \label{fig: generalSimPosAE}
\end{figure*}

Figure \ref{fig: generalSimPosAE} shows the positions and orbital elements of debris after ${440 \: {\rm Myr}}$ (the age of Fomalhaut) in one example simulation. For this run Fomalhaut b has a mass of ${0.1 \; {\rm M_{Jup}}}$ (${30 \; {\rm M_\oplus}}$), with a semimajor axis of ${203 \; {\rm au}}$ and eccentricity of ${0.760}$; this orbit corresponds to the minimum allowed eccentricity of Fomalhaut b if it orbits within ${5^\circ}$ of the disc midplane, as marked by a circle on Figure \ref{fig: fomalhautbElementsInRegionOfInterest}. 2000 debris particles were modelled in this simulation.

Most debris forms the V-shaped structure described by \citet{Beust2014}, which is a secular effect expected when particle orbits cross that of a highly eccentric perturber. These particles librate in eccentricity and apsidal orientation whilst their semimajor axes remain constant. The orbit of each particle periodically cycles between two extremes; it starts near-circular, then becomes highly eccentric (${e \sim 1}$) and misaligned by ${\sim 70^\circ}$ with the perturber orbit, before returning to the low-eccentricity case and continuing the cycle. The superposition of all such particles results in the V-shaped structure. These particles are visible on the right plot of Figure \ref{fig: generalSimPosAE} as the large population that has not undergone significant semimajor axis evolution, but which occupies the entire eccentricity range from 0 to very close to 1. Clearly the simulated secular structure does not resemble the narrow debris ring observed around Fomalhaut, and like \citet{Beust2014} we conclude that the observed ring cannot be in a long-term-stable configuration driven by a secular interaction with Fomalhaut b.

Another population visible on Figure \ref{fig: generalSimPosAE} is scattered debris. These particles are identifiable on the right plot as those whose semimajor axes differ considerably from their initial values. Since secular and resonant interactions do not significantly affect orbital energy, the semimajor axes of these particles must have changed through close encounters with the perturber. These debris particles undergo repeated scattering events, effectively performing a random walk in semimajor axis. After some time, the scattered particles are ultimately ejected from the system. This is the eventual fate of all secular debris; since secular evolution is unrelated to the position of the perturber on its orbit, ultimately each secular particle will make a close approach to the perturber and get scattered. Again, this was observed by \citet{Beust2014} who showed that the V-shaped secular structure eventually depletes as particles undergo repeated close encounters. The proportion of debris ejected over the system lifetime increases with planet mass, with only the most-massive ($1 \; \rm M_{Jup}$) simulated planets clearing a significant fraction of particles in our simulations. The scattered population forms a broad, diffuse cloud with little asymmetric structure, which again does not resemble the narrow debris ring observed around Fomalhaut.

At this point previous studies concluded that Fomalhaut b cannot have significant mass, because if it did then it would quickly destroy the observed debris ring through secular and/or scattering interactions. However, we find that in many simulations a third debris population exists: particles in mean-motion resonance with the perturber. These particles have semimajor axes close to the locations of nominal internal resonances (Equation \ref{eq: nominalResonanceLocation}). Like the secular population they oscillate in eccentricity, but in many cases their maximum eccentricity is much lower than in the secular case (${e \lesssim 0.3}$). Some resonant populations are also apsidally aligned with the orbit of the perturber. Unlike the secular and scattered populations, many resonant particles are stable for the system lifetime; they are protected from scattering because they never have a close encounter with the perturber, despite their orbits crossing. The orbits of some resonant particles bear a close resemblance to the debris ring observed around Fomalhaut. However, since the resonance widths (roughly, the range of particle semimajor axes occupying a resonance) are much narrower than the initial discs we consider, the numbers of resonant particles in these simulations are often too small to allow a detailed examination of resonant structures. In the following section we perform a different numerical strategy, allowing us to examine this long-term-stable debris without having to run prohibitively costly simulations.

\subsection{Numerical investigation of long-term-stable debris}
\label{subsec: numericalInvestigationStableOnly}

\noindent In the previous section we investigated the interaction between debris and a massive Fomalhaut b, and found that most particles in the scenario are unstable over the stellar lifetime. These particles undergo repeated scattering events, and many are eventually ejected from the system. However, we also identified several resonant populations that do appear stable. We now perform a detailed exploration of this long-term-stable debris.

\subsubsection{Setup of long-term-stable n-body simulations}
\label{subsec: numericalSetupStableOnly}

\noindent As before, we use {\sc rebound} n-body simulations to investigate the long-term stability of debris. The simulation setup is identical to that described in Section \ref{subsec: numericalSetupAllDebris}, and these simulations are again run for the lifetime of the star. The difference is that, in order to ensure a large population of stable particles whilst avoiding prohibitively long integration times, these simulations seek to remove unstable particles much sooner than the general simulations do.

The general simulations in Section \ref{subsec: numericalInvestigationAllDebris} showed that unstable particles undergo repeated scattering events, before eventually being ejected from the system. Additionally, initially unstable particles cannot become stable by being scattered into resonance, because they would eventually undergo another close encounter and scatter again. We can therefore identify unstable particles as those that are scattered at least once. This is done by considering particle semimajor axes; since secular and resonant interactions do not cause large semimajor axis changes, any particle whose semimajor axis differs significantly from its initial value must have been scattered. As these new simulations progress, they search for long-term-unstable debris by performing regular re-evaluations and identifying particles whose semimajor axes have changed by more than ${2 \; \rm per \; cent}$ since the start of the simulation (${2 \; \rm per \; cent}$ is larger than the semimajor axis libration experienced by our resonant particles). Any such particles are removed from the ongoing simulation. Since many unstable particles are removed in this way, these simulations can be run with more particles than the general simulations in Section \ref{subsec: numericalInvestigationAllDebris} (typically ${\sim 10^4}$ particles, rather than ${\sim 10^3}$). This allows a more detailed study of stable populations. However, care must be taken when interpreting the results, because any scattered population that would exist in reality is omitted. We run 60 of this type of simulation, with the same parameters as the general simulations in Section \ref{subsec: numericalInvestigationAllDebris} (Table \ref{tab: simulationsRun} in Appendix \ref{app: simulationsRun}).

\subsubsection{Results of long-term-stable n-body simulations}
\label{subsec: numericalResultsStableOnly}

Like the general simulations in Section \ref{subsec: numericalInvestigationAllDebris}, a significant fraction of debris in the new simulations undergoes non-negligible semimajor axis changes through close encounters with Fomalhaut b. Such particles are unstable over the stellar lifetime, and are therefore artificially removed from the new simulations. However, at the end of these simulations several debris populations remain, with semimajor axes very similar to their initial values. This debris is long-term stable, despite being on orbits that cross the simulated Fomalhaut b.

\begin{figure*}
  \centering
   \includegraphics[width=16cm]{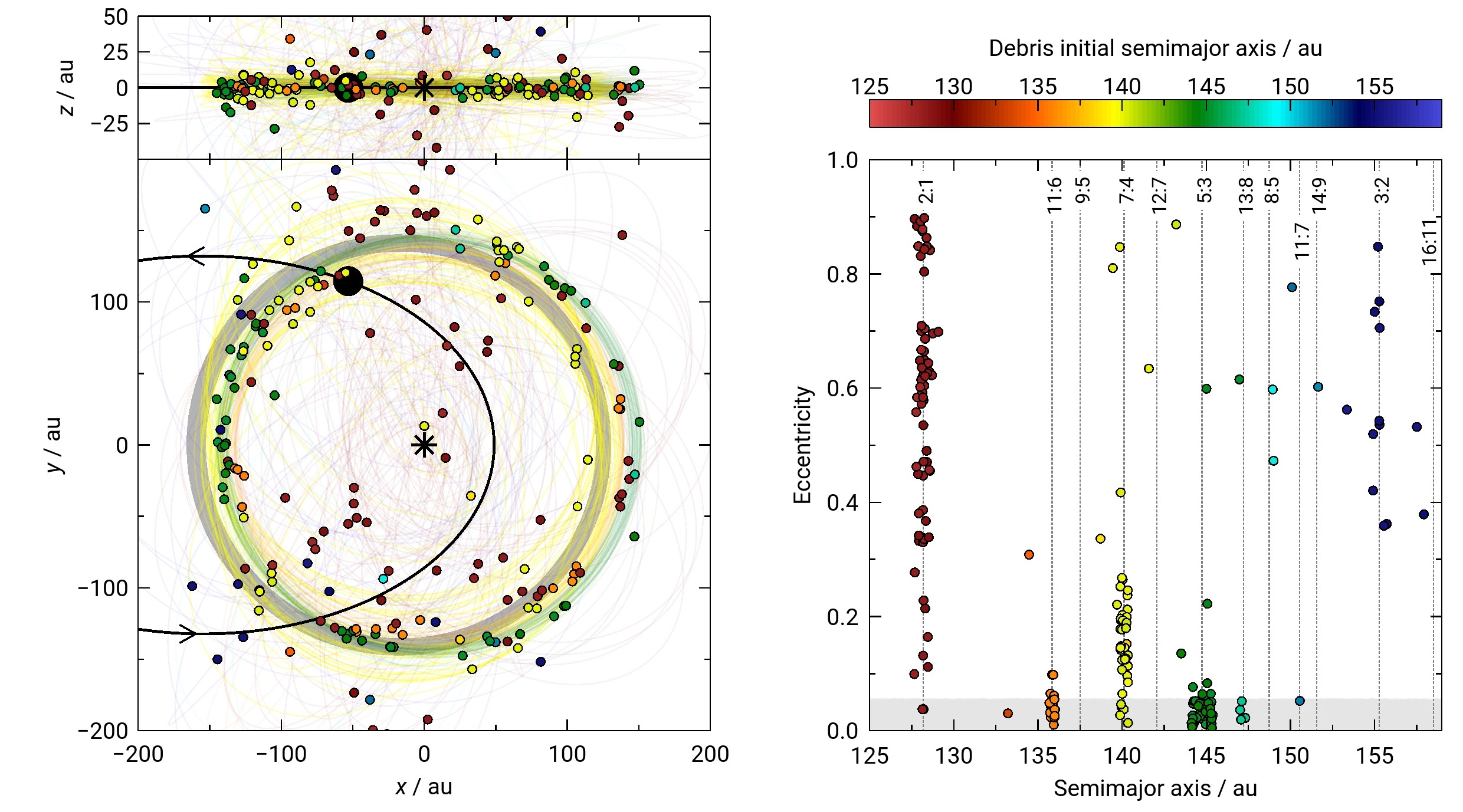}
   \caption{Long-term-stable debris in a simulation where unstable particles have been artificially removed, as described in Section \ref{subsec: numericalInvestigationStableOnly}. The simulation is shown at ${440 \; {\rm Myr}}$, the age of Fomalhaut, and has been further advanced so that Fomalhaut b (black circle on left plots) is at its 2004 location. Black arrows on the bottom-left plot show the direction of motion of Fomalhaut b. The simulation parameters are identical to the simulation on Figure \ref{fig: generalSimPosAE}, except that this simulation started with 40000 particles (rather than 2000 on Figure \ref{fig: generalSimPosAE}). The shaded grey region on the left plot is the deprojected observed disc, and thin lines show the instantaneous orbits of individual small bodies (coloured by their initial semimajor axis). Dotted lines on the right plot show the nominal locations of all internal mean-motion resonances of order ${q \leq 5}$; note that the semimajor axis range is that of the initial disc, and that the colour scheme differs from Figure \ref{fig: generalSimPosAE} to emphasise different debris populations. Whilst the majority of particles are unstable and have been artificially removed, several long-term-stable resonant populations exist. In particular, note particles in the 11:6, 7:4, 5:3 and 13:8 resonances on the right plot that inhabit stable, low-eccentricity orbits which cross that of the simulated Fomalhaut b. The left plots show that these particles (orange, yellow, green and turquoise points respectively) are on orbits similar to the observed disc.}
   \label{fig: removalSimsPosAvsE}
\end{figure*}

Figure \ref{fig: removalSimsPosAvsE} shows a simulation at ${440 \; {\rm Myr}}$, where particles whose semimajor axes have changed by more than ${2 \; \rm per \; cent}$ from their initial values have been removed. The simulation has the same initial parameters as that shown on Figure \ref{fig: generalSimPosAE}, except that the new simulation started with a much larger number of disc particles (${40000}$, as opposed to 2000 on Figure \ref{fig: generalSimPosAE}). The surviving debris exists in several distinct populations, defined by initial semimajor axes. We find that the average semimajor axis of each population coincides with the nominal location of a mean-motion resonance with the simulated Fomalhaut b (of order ${q \leq 5}$); this suggests that these particles are resonant, and that they survive because the resonance protects them from close encounters with the simulated planet. In Section \ref{subsec: evolutionOfResonantParticles} we will show that these stable particles are indeed in mean-motion resonances.

Whilst the results shown on Figure \ref{fig: removalSimsPosAvsE} are qualitatively similar for a broad range of simulation parameters, not all simulations have debris populations that are stable over the system lifetime. The parameter that has the greatest effect on whether any debris can be long-term stable appears to be the assumed mass of Fomalhaut b. Little or no long-term-stable particles exist in simulations with a ${1 \; {\rm M_{Jup}}} \; (300 \; {\rm M_\oplus})$ Fomalhaut b, but if the mass of the putative planet is of order ${0.1 \; {\rm M_{Jup}}} \; (30 \; {\rm M_\oplus})$ or lower then long-term-stable debris is almost always present. This is shown in Table \ref{tab: simulationsRun} in Appendix \ref{app: simulationsRun}, where ${f_{\rm stable}}$ denotes the fraction of initial particles whose semimajor axes change by less than ${2 \; \rm per \; cent}$ over the simulation; the value of ${f_{\rm stable}}$ strongly decreases as the mass of Fomalhaut b is increased. The reason for this could be because a greater degree of resonance overlap occurs with a more massive planet, causing particles to `bounce' between resonances and eventually get ejected (e.g. \citealt{Wisdom1980, Kuchner2003, Mustill2012}). It could also be because the larger the planet mass, the greater the semimajor axis change experienced by a small body passing close to it; the proportion of resonant particles on planet-crossing orbits that are destabilised by close encounters would therefore increase with planet mass. A combination of these effects likely precludes the long-term survival of debris if Fomalhaut b has a mass of order ${1 \; {\rm M_{Jup}}} \; (300 \; {\rm M_\oplus})$ or more. Conversely, if its mass is too low then few particles are destabilised through scattering events, and unexcited debris may survive for the stellar lifetime simply because the interaction timescale is too long. This is particularly true if the semimajor axis of Fomalhaut b is large, because this results in very long secular and scattering timescales. If Fomalhaut b does sculpt the observed debris disc, then we discuss the required mass of the putative planet in Section \ref{subsec: requiredSystemParameters}.

Other than mass, the remaining parameters of a coplanar Fomalhaut b (semimajor axis and eccentricity) do not strongly affect the existence of long-term-stable particles. For example, all simulations with a ${0.1 \; {\rm M_{Jup}}} \; (30 \; {\rm M_\oplus})$ Fomalhaut b have some particles whose semimajor axes change by less than ${2 \; \rm per \; cent}$ over the ${440 \; \rm Myr}$ simulation (${f_{\rm stable}}$ in Table \ref{tab: simulationsRun}). Furthermore, almost all of these simulations have particles that remain on unexcited, Fomalhaut disc-like orbits (${f_{\rm unexcited}}$ in Table \ref{tab: simulationsRun}: the fraction of initial disc particles that, in addition to not undergoing a semimajor axis change of more than ${2 \; \rm per \; cent}$, also never exceed an eccentricity of 0.4 or an inclination of ${20^\circ}$). The debris morphology and the number of stable, unexcited particles does vary slightly with the semimajor axis and eccentricity of Fomalhaut b, but in general the stable particle orbits are similar to those on Figure \ref{fig: removalSimsPosAvsE}. The existence of stable regions of parameter space is not limited  to a few specific perturber orbits, so it appears that stable, low-eccentricity, low-inclination orbits are possible for a wide range of highly eccentric perturbers, provided the mass of the latter is not too high. Note that regardless of the simulation parameters, only a small fraction of initial debris remains long-term stable, and an even smaller fraction remains on unexcited orbits; we discuss the implications of this in Section \ref{subsec: requiredSystemParameters}.

\subsubsection{Evolution of resonant particles}
\label{subsec: evolutionOfResonantParticles}

The simulations in Section \ref{subsec: numericalResultsStableOnly} show that some debris remains stable over the system lifetime, despite crossing the orbit of a highly eccentric, coplanar, massive perturber. In this section we show that these stable particles are in internal mean-motion resonances with the simulated Fomalhaut b, and investigate their evolution with a semi-analytic model.

Figure \ref{fig: removalSimsPosAvsE} shows that most long-term-stable particles have semimajor axes close to nominal resonances (Equation \ref{eq: nominalResonanceLocation}). However, this is not a sufficient condition for a  particle to be in resonance; to prove that these bodies are resonant, we must identify a resonant argument $\phi$ for them that librates around some angle, rather than circulating through ${360^\circ}$. A resonant argument depends on body positions through their mean longitudes

\begin{equation}
\lambda \equiv \varpi + \mathcal{M},
\label{eq: meanLongitude}
\end{equation}
 
\noindent where $\mathcal{M}$ is the mean anomaly. For particles in the ${p+q:p}$ internal resonance, the relevant $\phi$ value can be identified by considering disturbing function arguments containing ${\pm(p+q)\lambda_{\rm plt} \mp p \lambda}$ terms, where $\lambda_{\rm plt} $ and $\lambda$ are the mean longitudes of the simulated Fomalhaut b and particle respectively. Since we are interested in long-term-stable debris on low-eccentricity orbits that could resemble the Fomalhaut disc, the most interesting resonances shown on Figure \ref{fig: removalSimsPosAvsE} are the ${11{:}6}$, ${7{:}4}$, ${5{:}3}$ and ${13{:}8}$ internal resonances. Taking disturbing function arguments from \citet{Murray1999} and comparing them to our simulations, we find that the relevant resonant arguments for these four populations are

\begin{equation}
\begin{aligned}
\phi_{11{:}6} &= 11 \lambda_{\rm plt} - 6 \lambda - 5 \varpi_{\rm plt},\\
\phi_{7{:}4} &= 7 \lambda_{\rm plt} - 4 \lambda - 3 \varpi_{\rm plt},\\
\phi_{5{:}3} &= 5 \lambda_{\rm plt} - 3 \lambda - 2 \varpi_{\rm plt},\\
\phi_{13{:}8} &= 13 \lambda_{\rm plt} - 8 \lambda - 5 \varpi_{\rm plt}.
\end{aligned}
\label{eq: resonantArguments}
\end{equation}

\begin{figure}
  \centering
   \includegraphics[width=7cm]{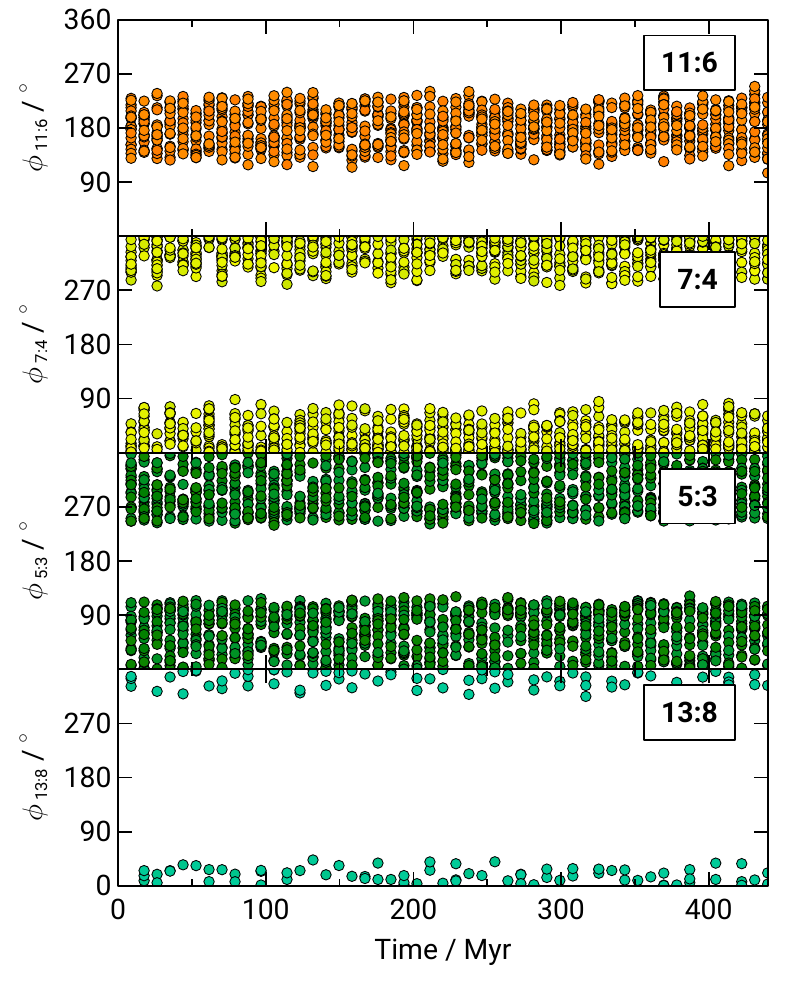}
   \caption{Resonant arguments of the four main low-eccentricity, long-term-stable debris populations from the simulation on Figure \ref{fig: removalSimsPosAvsE}, shown over the lifetime of the Fomalhaut system. The orbits of all particles cross that of the simulated Fomalhaut b. Particles are coloured by initial semimajor axis, using the colour scheme from Figure \ref{fig: removalSimsPosAvsE}. The resonant arguments librate rather than circulate, showing that all low-eccentricity, long-term-stable particles are in resonance with the simulated Fomalhaut b.}
   \label{fig: resonantArguments}
\end{figure}

\noindent We take the four low-eccentricity, long-term-stable debris populations that lie close to these resonances from Figure \ref{fig: removalSimsPosAvsE}, and plot their resonant arguments over the system lifetime on Figure \ref{fig: resonantArguments}. These resonant arguments clearly librate rather than circulate, and so we conclude that all simulated low-eccentricity particles that remain on stable, planet-crossing orbits for the system lifetime are indeed in internal resonances with the simulated Fomalhaut b.

We now investigate the expected evolution of resonant debris which crosses the orbit of a highly eccentric perturber, to better understand our simulation results. This extreme orbital configuration precludes the use of many theoretical resonance treatments, since these often assume a low-eccentricity perturber; however, we can estimate high-eccentricity resonant behaviour using a semi-analytical Hamiltonian approach, which has been well-utilised in the literature \citep{Yoshikawa1989, Beust1996, Beust2000, Faramaz2015, Faramaz2017, Pichierri2017}. We refer the reader to \cite{Beust2016} for a detailed description of this method, which we only briefly outline here.

As described above, resonant motion is characterised by the libration of a resonant argument (e.g. Equation \ref{eq: resonantArguments}). For a given resonant particle, its averaged Hamiltonian is obtained by time-averaging its instantaneous Hamiltonian over the perturber motion, assuming a fixed resonant argument. For coplanar bodies, the result is a conservative Hamiltonian with two degrees of freedom. If the perturber has zero eccentricity, then another constant of motion appears and the Hamiltonian becomes integrable \citep{Moons1995}; in this case the resonant argument librates around an equilibrium. For eccentric perturbers like ours the motion is more complex, but to lowest order the libration and its amplitude are preserved \citep{Morbidelli1993}. Furthermore, if one assumes initially zero-amplitude librations then the Hamiltonian reduces to one degree of freedom, even for non-zero perturber eccentricities. This technique remains valid for high planet eccentricities \citep{Pichierri2017}, and we use this method to determine the Hamiltonians of resonant particles.

\begin{figure*}
  \centering
   \includegraphics[width=16cm]{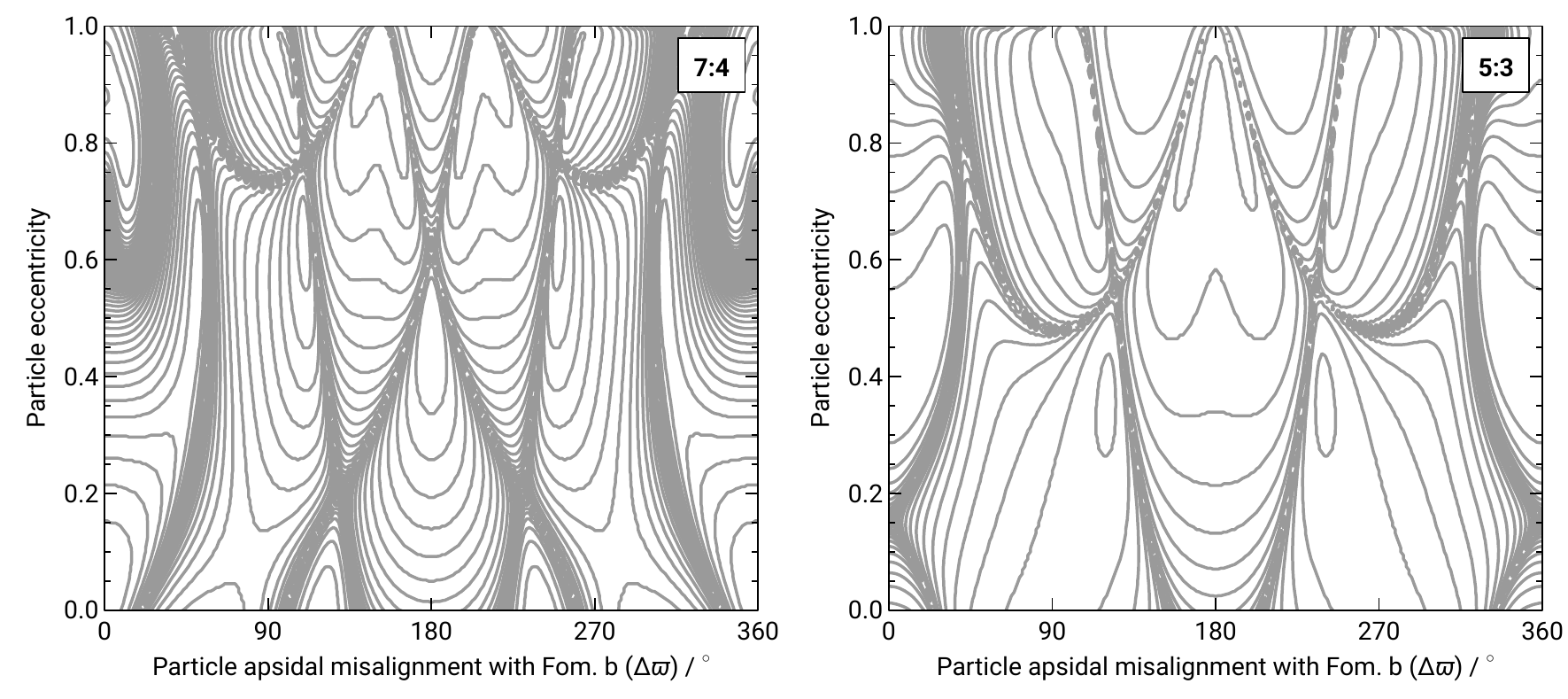}
   \caption{Theoretical Hamiltonians of particles in the 7:4 (left) and 5:3 (right) internal resonances with a coplanar Fomalhaut b, if the latter has semimajor axis ${203 \; \rm au}$, eccentricity 0.760 and mass ${0.1 \; \rm M_{Jup}}$ (${30 \; {\rm M_\oplus}}$). A resonant particle placed on one of these contours would move around that contour, undergoing coupled librations in eccentricity and apsidal orientation as described in Section \ref{subsec: evolutionOfResonantParticles}. Most of the low-eccentricity resonant particles in the 7:4 and 5:3 resonances on Figure \ref{fig: removalSimsPosAvsE} occupy the low-eccentricity libration islands around ${\Delta \varpi = 0}$ on this plot, meaning that their eccentricities are maximised when apsidally aligned with the orbit of Fomalhaut b (although some 7:4 particles also occupy misaligned libration regions, giving rise to the two misaligned `lobes' visible in the yellow orbits on Figure \ref{fig: removalSimsPosAvsE}). Whilst some particles could theoretically occupy the various high-eccentricity libration islands also visible on this plot, our n-body simulations show that these high-eccentricity modes are not long-term stable.}
   \label{fig: 5_3And7_4PhaseMaps}
\end{figure*}

On Figure \ref{fig: 5_3And7_4PhaseMaps} we plot such theoretical Hamiltonian contours for particles in the internal 7:4 and 5:3 resonances with the simulated Fomalhaut b, assuming that the latter has a semimajor axis, eccentricity and mass of ${203 \; \rm au}$, 0.760 and ${0.1 \; \rm M_{Jup}}$ (${30 \; \rm M_\oplus}$) respectively (the same parameters as Fomalhaut b on Figures \ref{fig: generalSimPosAE}-\ref{fig: resonantArguments}). The contours are plotted as functions of particle eccentricity and pericentre orientation relative to that of the planet (${\Delta \varpi \equiv \varpi - \varpi_{\rm plt}}$). If a particle has a resonant libration amplitude of zero, and is located on one of the plotted contours, then it will only move along that contour. The lines on Figure \ref{fig: 5_3And7_4PhaseMaps} therefore show the coupled libration in eccentricity and pericentre orientation expected of a particle with zero libration amplitude in resonance with the simulated Fomalhaut b; particles with small resonant librations would experience small additional oscillations around these curves \citep{Beust2000}.

Using Figure \ref{fig: 5_3And7_4PhaseMaps}, one can understand the behaviour of resonant particles in our n-body simulations. Given the initial disc parameters described in Section \ref{subsec: numericalSetupAllDebris}, at the start of the simulations all particles occupy a narrow band across the bottom of the Figure \ref{fig: 5_3And7_4PhaseMaps} plots, with eccentricities between 0 and 0.05 and pericentre orientations spanning the full ${0-360^\circ}$ range. This means that resonant debris initially occupies several distinct libration regions. For the 7:4 resonance, one such libration island is centred on ${\Delta \varpi = 0}$, ${e \approx 0.1}$; the plot shows that debris in this resonance with initial ${|\Delta \varpi| \lesssim 25^\circ}$ will librate around this central point, with eccentricity oscillating between 0 and up to 0.3 and ${\Delta \varpi}$ oscillating between ${\pm 40^\circ}$. This matches most of the low-eccentricity 7:4 resonant population in our simulations (yellow points on Figure \ref{fig: removalSimsPosAvsE}), although some 7:4 particles also occupy misaligned libration regions resulting in two misaligned `lobes' visible in the yellow orbits on Figure \ref{fig: removalSimsPosAvsE}. A low-eccentricity population is also predicted for the 5:3 resonance, as shown on the right plot of Figure \ref{fig: 5_3And7_4PhaseMaps}; these particles are predicted to have eccentricities smaller than ${\sim 0.1}$, and again this population is visible in our n-body simulations (green points on Figure \ref{fig: removalSimsPosAvsE}). Figure \ref{fig: 5_3And7_4PhaseMaps} also predicts the existence of highly eccentric, misaligned resonant debris populations, most notably around ${\Delta \varpi = 180^\circ}$; however, these particles do not appear long-term stable in our n-body simulations, with very few such bodies visible on Figure \ref{fig: removalSimsPosAvsE}. The reason for this is unclear, but it seems that stable particles in the 11:6, 7:4, 5:3 and 13:8 resonances favour low-eccentricity libration modes. The eccentricities are not too dissimilar to that of the Fomalhaut disc (${e_{\rm disc} \approx 0.12}$), so these resonances are of greatest interest for this paper.

The results in this section demonstrate that debris particles on orbits similar to the Fomalhaut disc can be long-term stable despite crossing the orbit of a massive, coplanar Fomalhaut b, provided that they occupy internal mean-motion resonances. This means that the argument that Fomalhaut b cannot have significant mass without disrupting the observed disc is not necessarily correct. However, whilst resonant particle \textit{orbits} are similar to the observed disc, the \textit{locations} of particles on those orbits clearly differ from observations; our simulated resonant particles have a broad, clumpy morphology (Figure \ref{fig: removalSimsPosAvsE}), whilst the imaged ring is narrow and smooth. In the following section we show that, provided at least one inner planet is also present in the system, the millimetre dust produced by collisions between resonant bodies can resemble the observed debris disc.

\section{The Fomalhaut disc as a resonant ring}
\label{sec: fomalhautDiscAsResRing}

We have shown that resonant debris on low-eccentricity, apsidally aligned orbits similar to the Fomalhaut disc is stable for the system lifetime, despite crossing the orbit of a coplanar, $0.1 \; \rm M_{Jup}$ (${30 \; {\rm M_\oplus}}$) Fomalhaut b. This means that Fomalhaut b can be massive and coplanar without disrupting the disc. However, whilst the simulated debris orbits are similar to the observed disc, the distribution of particles on those orbits differs from the narrow, smooth ring observed at millimetre and shorter wavelengths. In this section we show that, if we include collisions and at least one additional inner planet, then our resonant model can better reproduce the Fomalhaut debris disc.
\subsection{Removing high-eccentricity debris through interactions with the inner system}
\label{subsec: removingUnstableDebris}

We have shown that debris in resonance with a $0.1 \; \rm M_{Jup}$ (${30 \; {\rm M_\oplus}}$) Fomalhaut b can be long-term stable on orbits similar to the observed disc, whilst non-resonant debris is unstable and undergoes repeated scattering encounters with the putative planet. However, Figure \ref{fig: generalSimPosAE} shows that most unstable debris is not ejected within the stellar lifetime, and so the overall debris morphology therefore does not resemble the observed narrow ring. What the simulated Fomalhaut b \textit{can} do though is periodically drive some particle eccentricities up close to 1 through secular interactions (Section \ref{subsec: numericalResultsAllDebris}). Since the semimajor axes of secular and resonant particles are roughly constant, this eccentricity increase drives particle pericentres right down into the inner system (${q \lesssim 1 \; \rm au}$). Any planets in these inner regions could potentially scatter and eject this debris, leaving only the low-eccentricity resonant ring. The existence of inner planets is not an unreasonable suggestion, because if Fomalhaut b is massive then a dynamical interaction with another body may have been required to drive it onto its eccentric orbit (e.g. \citealt{Kalas2013, Faramaz2015}); for example, one possibility is that Fomalhaut b formed in the inner system, got scattered outwards by a large planet, then decoupled from that planet due to interactions with debris (\citealt{Pearce2014} show that the pericentre distance of the eccentric Fomalhaut b would increase, protecting it from further scattering by the inner planet).

To test whether inner planets could eject high-eccentricity debris, we run another simulation with the same setup as that on Figure \ref{fig: generalSimPosAE}, but this time also including a hypothetical ${1 \rm \; M_{Jup}}$ planet on a circular orbit at ${10 \; \rm au}$. This planet is allowed by direct imaging \citep{Kenworthy2009, Kenworthy2013}. The only other difference between this simulation setup and that on Figure \ref{fig: generalSimPosAE} is that more particles are used: 40,000 rather than 2000. Particles are only removed from the simulation if their stellocentric distance exceeds ${10^4 \; {\rm au}}$, i.e. unstable particles are \textit{not} artificially removed (unlike the simulation on Figure \ref{fig: removalSimsPosAvsE}).

\begin{figure*}
  \centering
   \includegraphics[width=16cm]{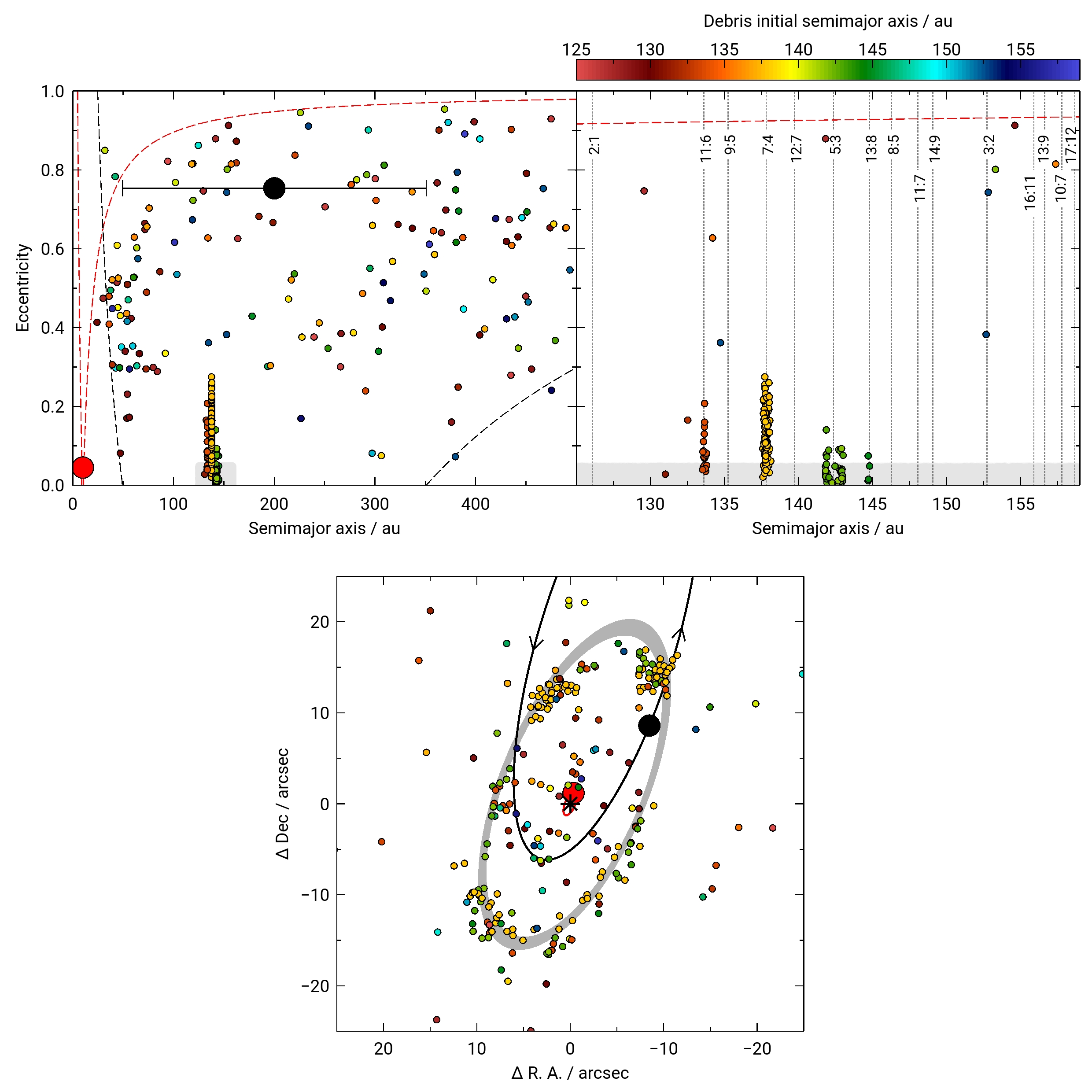}
   \caption{Simulation with a hypothetical ${1 \rm \; M_{Jup}}$ planet orbiting at ${10 \; \rm au}$ (red circle and lines), in addition to a ${0.1 \rm \; M_{Jup}}$ (${30 \; {\rm M_\oplus}}$) Fomalhaut b; the inner planet has removed almost all unstable and high-eccentricity debris, leaving resonant debris on orbits similar to the Fomalhaut ring. The simulation setup is described in Section \ref{subsec: removingUnstableDebris}. Particles were only removed if their stellocentric distance exceeded ${10^4 \; {\rm au}}$, i.e. unstable particles were \textit{not} artificially removed (unlike the simulation on Figure \ref{fig: removalSimsPosAvsE}). The remaining simulation parameters are identical to the simulations on Figures \ref{fig: generalSimPosAE} and \ref{fig: removalSimsPosAvsE}, except that this simulation started with 40000 particles (rather than 2000 on Figure \ref{fig: generalSimPosAE}). The simulation is shown at ${440 \; {\rm Myr}}$, and has been further advanced so that Fomalhaut b is at its 2004 location. The top plots show semimajor axes and eccentricities, with the range of semimajor axes on the top-right plot equal to the range of semimajor axes of the initial disc. Particles between the pairs of dashed black or red lines may cross the orbits of Fomalhaut b or the inner planet, respectively. The bottom plot shows particle positions, rotated to the orientation of the Fomalhaut disc on the sky. North is up and east is left, and the shaded grey region on the bottom plot is the observed debris ring. All other symbols are the same as previous plots. Note that the small interaction between Fomalhaut b and the hypothetical inner planet has caused the nominal resonance locations to shift slightly compared to Figure \ref{fig: removalSimsPosAvsE}.}
   \label{fig: fomcSimSkyPosAVsE}
\end{figure*}

Figure \ref{fig: fomcSimSkyPosAVsE} shows this simulation after the stellar lifetime of ${440 \; \rm Myr}$. The inclusion of a hypothetical inner planet has clearly altered the system evolution, and very little unstable debris now survives. This is in marked contrast to simulations without an inner planet (Figure \ref{fig: generalSimPosAE}), where a significant quantity of unstable debris remains. We discussed the reason above; the secular effect of Fomalhaut b drives the pericentres of most debris particles down into the inner system, where they now encounter the hypothetical inner planet. This planet is much more efficient at ejecting debris than the simulated Fomalhaut b is, owing to its larger mass, smaller semimajor axis and lower eccentricity, and it clears most non-resonant debris within the system lifetime. The V-shaped secular structure and diffuse scattered population on Figure \ref{fig: generalSimPosAE} are therefore greatly reduced by the inclusion of an inner planet.

Not only does an inner planet clear unstable debris, but it also removes some stable resonant particles whose orbits do not resemble the observed disc. Debris morphologies in simulations with an inner planet (Figure \ref{fig: fomcSimSkyPosAVsE}) therefore more closely resemble the observed disc than the morphologies in simulations without an inner planet, but where unstable particles has been artificially removed (Figure \ref{fig: removalSimsPosAvsE}). This is because simulations without an inner planet host some highly eccentric resonant debris (most notably that in the 2:1 resonance on Figure \ref{fig: removalSimsPosAvsE}), which is protected from close encounters with Fomalhaut b and is therefore long-term stable. Such debris does not occupy low-eccentricity, apsidally aligned orbits, and therefore looks very different to the Fomalhaut disc. By including planets in the inner system, we remove all secular and high-eccentricity resonant debris, leaving only low-eccentricity, apsidally aligned populations in the 11:6, 7:4, 5:3 and 13:8 resonances. These particles are remarkably unscathed, since they remain on low-eccentricity orbits and so never encounter the inner system. This means that resonances protect such particles twice: firstly from encounters with Fomalhaut b, and secondly from any planets in the inner system.

Whilst we included a hypothetical ${1 \rm \; M_{Jup}}$ planet at ${10 \; \rm au}$ in this simulation, this is merely an example inner planet and not a prediction of specific planet parameters; many different planet configurations would have the desired effect of ejecting debris driven into the inner system by Fomalhaut b. The only real constraint is that any major inner planets should be interior to ${\sim 50 \; \rm au}$ in this specific simulation, otherwise they would interact with Fomalhaut b at pericentre and could cause the orbit of the latter to rapidly evolve, which may destabilise resonant debris. Even this is not a general constraint, because if Fomalhaut b lies within $5^\circ$ of the disc plane then its pericentre could be as large as ${65 \; \rm au}$ (Figure \ref{fig: fomalhautbAllowedOrbits}). Regarding the mass of any inner planets, for this specific simulation we find that a single ${1 \rm \; M_{Jup}}$ planet at ${10 \; \rm au}$ is sufficient to clear debris within the stellar lifetime, whilst a ${0.1 \rm \; M_{Jup}}$ (${30 \; {\rm M_\oplus}}$) planet at this location is not. However, this again does not constitute a lower limit on any planets in the inner system, since the required mass will differ at different orbital locations. Multiple, lower-mass planets could also clear debris more effectively than a single, higher-mass planet could, so any planet constraints would depend on the number of inner planets assumed to exist.
 
In this section we showed that planets in the inner Fomalhaut system could remove unstable and high-eccentricity debris within the system lifetime, leaving only resonant material on low-eccentricity, apsidally aligned orbits that resemble the observed ring. Whilst these resonant orbits broadly match the imaged disc, the simulated parent body population is not as smooth or as narrow as that in observations. However, we now show that dust produced in collisions between resonant bodies is expected to better match the observed debris ring.

\subsection{Azimuthal gaps, clumps, and ring width: the role of collisions between resonant bodies}
\label{subsec: gapsAndCollisions}

\noindent Figure \ref{fig: fomcSimSkyPosAVsE} shows that the simulated resonant ring is not narrow, smooth and continuous, but rather broad with clumps and azimuthal gaps. This is in contrast to ALMA observations, which show the Fomalhaut disc to be azimuthally smooth at ${1.3 \; \rm mm}$ \citep{White2017, Macgregor2017}. The observed ring does have an azimuthal gap in visible light at position angle $331^\circ$ \citep{Kalas2013}, but this underdensity is much narrower than the broad depleted region in our simulations near that location (Figure \ref{fig: fomcSimSkyPosAVsE}). However, our simulations model large \textit{parent bodies}, whilst observations show millimetre or smaller \textit{dust}. In this section we show that dust produced by collisions in the resonant ring is expected to have a much smoother and narrower distribution than that of the parent bodies, and that this dust better reproduces the observed Fomalhaut disc.

\subsubsection{Distribution of resonant parent bodies}
\label{subsec: distributionOfResonantParentBodies}

\noindent The simulated resonant parent bodies have a broad, non-axisymmetric distribution. Non-axisymmetric features are an inevitable property of resonant populations, and are observed in all our simulations with resonant bodies. Clumps are a well-known resonant feature, arising from the `looping' trajectories of resonant bodies in a reference frame rotating with the perturber (e.g. \citealt{Ozernoy2000, Wyatt2002, Wyatt2006}). Two prominent clumps are visible in the northern half of the simulated ring on Figure \ref{fig: fomcSimSkyPosAVsE}. In addition to clumps, azimuthal gaps are also present; two such gaps can be seen on Figure \ref{fig: fomcSimSkyPosAVsE}, one at the far north of the simulated ring, and one near Fomalhaut b. These azimuthal gaps are not merely underdensities between clumps, but rather regions where resonant debris would coincide with the location of Fomalhaut b as it crosses the ring; any such debris is ejected, leaving gaps \citep{Pearce2014}. Clumps and azimuthal gaps are therefore an unavoidable feature of debris in long-term resonance with a disc-crossing perturber. 

\begin{figure*}
  \centering
   \includegraphics[width=16cm]{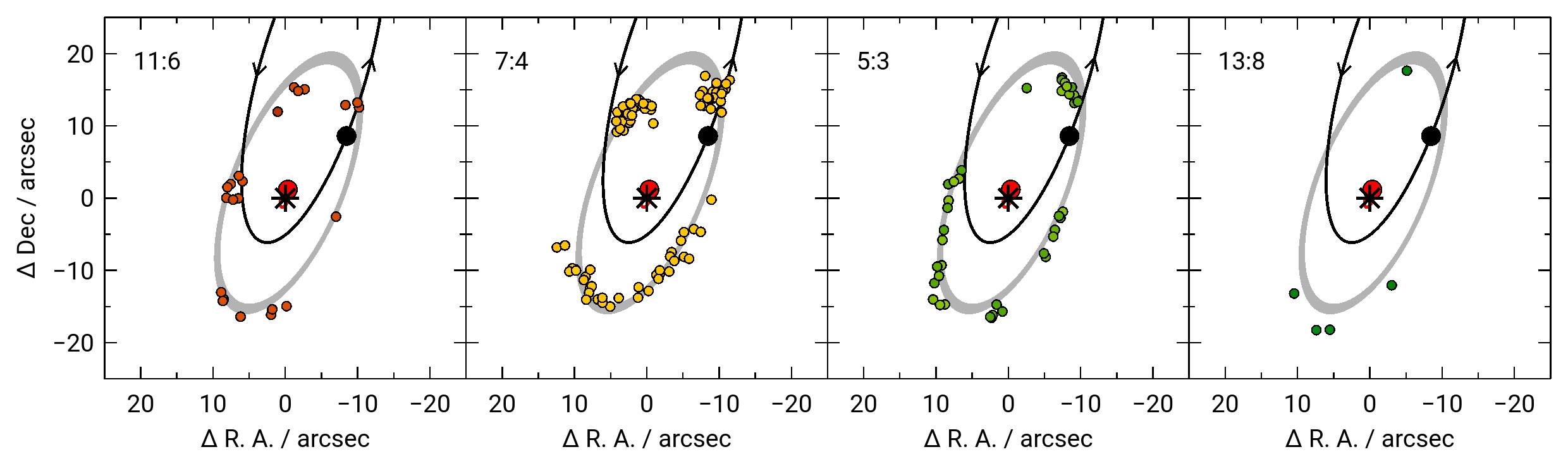}
   \caption{Resonant parent bodies from Figure \ref{fig: fomcSimSkyPosAVsE}, showing the different morphologies associated with different resonances. Note the different semimajor axes, eccentricities, orientations and clump sizes and positions. The superposition of these four populations results in a parent body distribution that is broader than the observed Fomalhaut ring (grey region).}
   \label{fig: fomcSkyPosSepRes}
\end{figure*}

In addition to non-axisymmetric features, resonant debris would also have a non-negligible radial extent. There are two reasons for this. Firstly, the `looping' trajectories of resonant bodies described above can produce clumps that are relatively wide, even if the particles in that resonance span only a narrow semimajor axis range. Secondly, our simulated debris ring is not a single resonant population at one semimajor axis, but rather a superposition of four different resonant populations. These populations have different semimajor axes, eccentricities, apsidal alignments, clump locations and widths, and so each individual resonant population has a different morphology. These different morphologies are shown on Figure \ref{fig: fomcSkyPosSepRes}; the superposition of these four resonant populations creates a ring that is broader than the observed disc.

If the Fomalhaut ring is a resonant remnant of an initially broad disc, then any long-term-stable debris must have the clumpy, gapped structure described above. These particles are the large parent bodies in our simulations, which survive for the stellar lifetime by occupying resonances protecting them from Fomalhaut b. Conversely, the dust observed in the Fomalhaut system would be much younger, and would have been recently released through collisions between resonant parent bodies. We now predict the distribution of this dust, to determine how it compares to the observed debris disc.

\subsubsection{Dust released in a single collision}
\label{subsec: dustFromSingleCollision}

\noindent Dust released in a collision between resonant bodies is not necessarily resonant itself, for several reasons. Firstly, a collision would release dust at a range of velocities, and hence a range of orbits. Even a ${1 \; \rm per \; cent}$ shift in semimajor axis is sufficient to move a body out of resonance (Figure \ref{fig: removalSimsPosAvsE}); dust released with only a modest velocity distribution could have such a semimajor axis spread, and may therefore be non-resonant \citep{Krivov2007}. Secondly, the colliding parents may belong to two different resonances, so the semimajor axes of ejected dust need not correspond to either nominal resonance location. Finally, sub-millimetre dust would be affected by radiation forces, so a small grain released with exactly the velocity of a resonant parent would still occupy a different, non-resonant orbit \citep{Wyatt2006}. Since dust orbits would therefore differ from those of their parents, the dust morphology would differ from the resonant parent bodies.

We first explore the morphology of millimetre dust released in a single example collision between two resonant parent bodies. In the simulation on Figure \ref{fig: fomcSimSkyPosAVsE}, at ${440 \; \rm Myr}$ there are two debris particles in the 7:4 resonance within ${2 \; \rm au}$ of each other, with a relative velocity of ${\sim 100 \; \rm m \; s^{-1}}$. We consider dust released in a hypothetical catastrophic collision between these two parents. This dust cloud would carry the sum of momenta of both colliders \citep{Krivov2006}, and would expand and eventually disperse owing to a spread of initial ejecta velocities. We model this dispersal by inserting 1000 particles representing dust into the simulation at the collision time. These particles are initially placed at the mean location of the two colliders, with mean velocity equal to the mean velocity of the colliders. To simulate a velocity dispersion, each dust grain is given a small additional random velocity component, with uniform random orientation and with a magnitude drawn uniformly between 0 and some maximal value ${\Delta v}$. We use a velocity dispersion of ${\Delta v = 5 \rm \; m \; s^{-1}}$, which is conservative given the collision velocity of ${\sim 100 \; \rm m \; s^{-1}}$; we discuss the implications of this later in this section. We assume the dust to be millimetre grains for comparison with ALMA data, and therefore omit radiation forces. After these bodies are inserted the simulation is continued, so as dust disperses it is also perturbed by Fomalhaut b and the inner planet. Further collisions within this dust population are not directly simulated, but are discussed below. 

\begin{figure*}
  \centering
   \includegraphics[width=16cm]{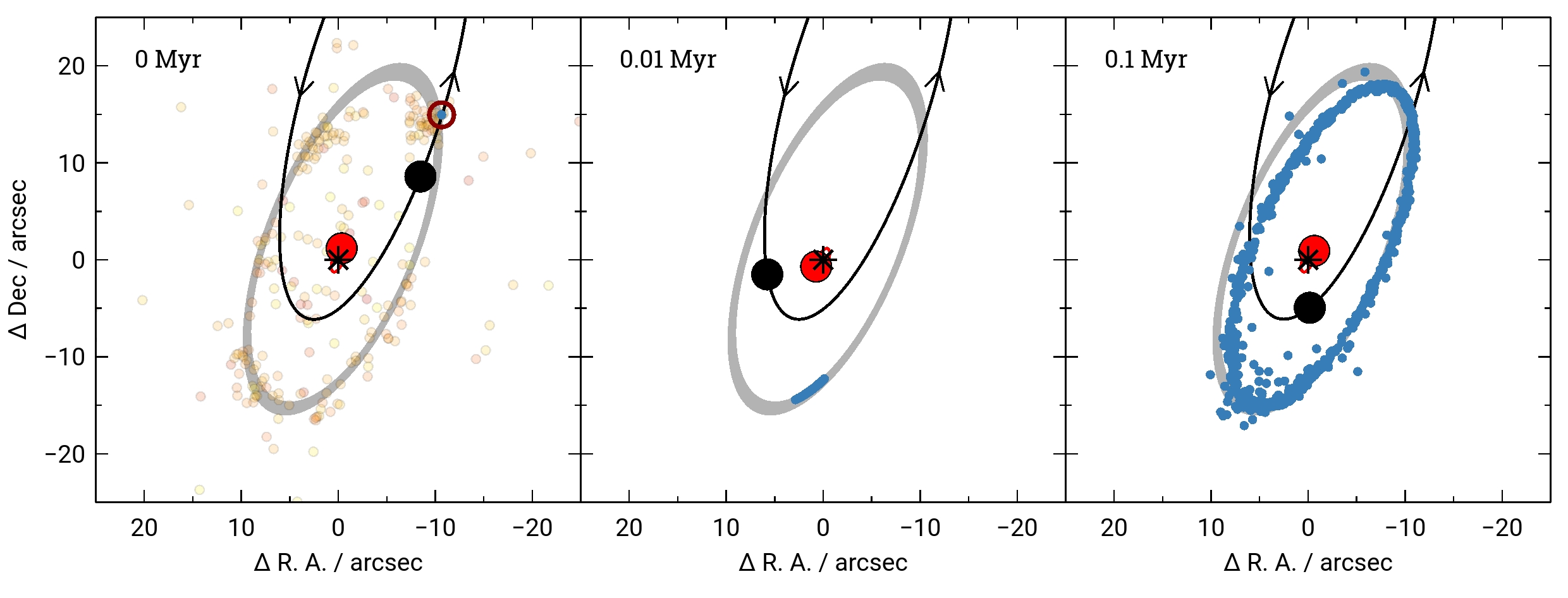}
   \caption{Simulation of millimetre dust released in one particular hypothetical collision between two closely approaching bodies from the simulation on Figure \ref{fig: fomcSimSkyPosAVsE}, as described in Section \ref{subsec: dustFromSingleCollision}. The blue points are dust, the transparent orange points are the parent bodies from Figure \ref{fig: fomcSimSkyPosAVsE}, and other symbols are defined on previous figures. At the time when the two example parent bodies make a close approach (${440 \; \rm Myr}$), 1000 dust particles are inserted into the simulation at the close approach location (brown ring on leftmost plot), with the mean velocity of the two parent bodies and an additional small velocity dispersion of ${5 \; \rm m \; s^{-1}}$. The simulation is then continued (including perturbations from Fomalhaut b and an inner planet), with the time since collision shown in the top-left of each plot. The figure shows that collisional ejecta shear into an azimuthally smooth distribution long before a ${0.1 \; {\rm M_{Jup}}}$ (${30 \; \rm M_\oplus}$) Fomalhaut b would significantly perturb their orbits. In particular, note the absence of the clumps and azimuthal gaps seen in the parent body population (Figure \ref{fig: fomcSimSkyPosAVsE}). An important point is that the shape and orientation of this specific dust ring is set by the positions and velocities of the two parent bodies at the moment of collision; collisions between different parent bodies would produce different dust rings, with different shapes and degrees of smoothness. In Section \ref{subsec: dustFromMultipleCollisions} and on Figure \ref{fig: collisionsHeatmap} we show that the superposition of dust from many such collisions well-reproduces the Fomalhaut disc.}
   \label{fig: singleCollisionPositions}
\end{figure*}

The simulation of dust dispersal from this particular collision is shown on Figure \ref{fig: singleCollisionPositions}; note that collisions between different parent bodies would produce different dust distributions, as discussed Section \ref{subsec: dustFromMultipleCollisions}. Despite perturbations from Fomalhaut b and an inner planet, the dust from this single collision has sheared out into a narrow, azimuthally smooth ring, similar to that observed. This is because shearing occurs significantly faster than a  ${0.1 \; {\rm M_{Jup}}}$ (${30 \; \rm M_\oplus}$) Fomalhaut b perturbs the ring, despite Fomalhaut b repeatedly passing through the dust (it crosses the ring 100 times during the simulation). The majority of simulated dust is not resonant, so over a long time the simulated particles would move out of a neat ring and form a structure like that on Figure \ref{fig: generalSimPosAE}, before eventually being ejected by scattering. However, in reality grains may never evolve out of the neat ring before being destroyed in further collisions; in order to determine the true fate of dust, we must therefore consider the timescales of the various physical processes.

The three timescales that determine the morphology of dust released in collisions are the time it takes dust to shear into a ring, the time it takes Fomalhaut b to perturb the dust, and the time it takes for grains to be destroyed through collisions. Firstly, the shearing time; if dust is released from a body on a circular orbit at radius $r$, and that dust is not significantly affected by radiation pressure, then the dust will shear into a ring over a timescale

\begin{equation}
t_{\rm shear} \sim \frac{r}{\Delta v},
\label{eq: shearingTime}
\end{equation}

\noindent where ${\Delta v}$ is the initial dust velocity dispersion (Equation \ref{eq: shearingTime} is derived in Appendix \ref{app: shearingTimeDerivation}). For the collision shown on Figure \ref{fig: singleCollisionPositions}, Equation \ref{eq: shearingTime} predicts that dust will shear into a smooth ring after ${t_{\rm shear} \sim 10^5 \; \rm yr}$, in agreement with the simulation. The second timescale is that for Fomalhaut b to perturb debris, which is roughly the secular timescale from second-order secular theory: 

\begin{equation}
t_{\rm sec} \approx 4 T_{\rm plt} \left(\frac{M_{\rm plt}}{M_*}\right)^{-1} \alpha^{-1/2} \left[ b^{(1)}_{3/2}(\alpha)\right]^{-1},
\label{eq: secularTimescale}
\end{equation}

\noindent where $T_{\rm plt}$ is the orbital period of Fomalhaut b, $\alpha \equiv a/a_{\rm plt}$, and 

\begin{equation}
b^{(j)}_{s}(\alpha) \equiv \frac{1}{\upi} \int^{2\upi}_{0} \frac{\cos(j \psi)}{(1 - 2 \alpha \cos \psi + \alpha^2)^s} {\rm d}\psi
\label{eq: leplaceCoeff}
\end{equation}

\noindent is a Laplace coefficient\footnote{Note that Equation \ref{eq: secularTimescale} has a different $\alpha$ dependence to Equation 17 in \citet{Pearce2014}; the former is for external perturbers (${a_{\rm plt}>a}$), whilst the latter is for internal perturbers (${a_{\rm plt}<a}$).} \citep{Murray1999}. For the dust in the simulation on Figure \ref{fig: singleCollisionPositions}, ${\alpha \approx 0.70}$, ${b^{(1)}_{3/2}(\alpha) \approx 7.5}$ and Equation \ref{eq: secularTimescale} predicts that Fomalhaut b would perturb the dust over ${t_{\rm sec} \sim 10^7 \; \rm yr}$; this is much longer than the shearing time (${t_{\rm shear} \sim 10^5 \; \rm yr}$), so dust released in a collision would shear into a ring before a ${0.1 \; {\rm M_{Jup}}}$ (${30 \; \rm M_\oplus}$) Fomalhaut b could induce azimuthal asymmetries. Finally, the third timescale is that for millimetre dust to be collisionally destroyed. We roughly estimate this by modelling collisions with similar-sized grains in a narrow ring, using the method of \cite{Wyatt2002}. In this case the collisional lifetime of dust is

\begin{equation}
t_{\rm col} \approx  \frac{8 \upi \rho_{\rm d} s_{\rm d} r^{2.5} \Delta r}{9 M_{\rm d} \sqrt{G M_*}},
\label{eq: collisionTimescale}
\end{equation}

\noindent where ${\rho_{\rm d}}$, $s_{\rm d}$, ${\Delta r}$, ${M_{\rm d}}$ and $G$ are the dust grain density, grain radius, disc width, total mass in millimetre dust, and the gravitational constant, respectively. Note we have assumed the dust inclinations to be comparable to their eccentricities. \cite{Macgregor2017} estimate the millimetre dust mass of the Fomalhaut ring to be ${M_{\rm d} = 0.015 \pm 0.010 \; \rm M_\oplus}$, and taking ${\rho_{\rm d} \approx 2 \; \rm g \; cm^{-3}}$ yields ${t_{\rm col} \sim 10^6 \; \rm yr}$. This is longer than the shearing time (${t_{\rm shear} \sim 10^5 \; \rm yr}$), but shorter than the secular time (${t_{\rm sec} \sim 10^7 \; \rm yr}$).

Taking the three timescales together, we can estimate the evolution and morphology of dust released in a single collision between resonant bodies. Dust would be released in a clump, which would then shear out into a narrow ring as shown on Figure \ref{fig: singleCollisionPositions}. This ring would be azimuthally smooth, since the shearing occurs much more rapidly than Fomalhaut b perturbs the dust. Dust would then collisionally deplete, and the ring would fade before Fomalhaut b could introduce significant azimuthal structure. This means that, aside from dust released in recent collisions, the Fomalhaut ring is expected to appear azimuthally smooth even if the dust is released by collisions between clumpy resonant bodies. We will show this in Section \ref{subsec: dustFromMultipleCollisions}. Note that a more sophisticated collision treatment may yield a different grain lifetime; however, provided that lifetime is at least comparable to the shearing time and less than the secular time, the qualitative results here are expected to hold.

The above results assume that the parent bodies in the Fomalhaut ring are resonant with a ${0.1 \; {\rm M_{Jup}}}$ (${30 \; \rm M_\oplus}$), coplanar Fomalhaut b with semimajor axis ${203 \; \rm au}$ and eccentricity 0.760. Different parameters for Fomalhaut b would change the interaction timescale, which would affect these conclusions. In particular, if Fomalhaut b had a mass of ${1 \; {\rm M_{Jup}}}$ (${300 \; \rm M_\oplus}$) then the secular timescale would be ${t_{\rm sec} \sim 10^6 \; \rm yr}$, which is similar to the collisional timescale, and the dust ring would show significant azimuthal structure before collisionally depleting. Since our previous simulations showed that resonant debris is not long-term stable if Fomalhaut b has such a high mass (Section \ref{subsec: numericalResultsStableOnly}), this further demonstrates that our scenario probably requires Fomalhaut b to be less massive than Jupiter to reproduce the Fomalhaut disc.

The results also depend on the dust initial velocity dispersion ${\Delta v}$, which sets the rate that dust shears out into a ring. This is roughly the speed that ejecta are released following a collision, ${v_{\rm ej}}$. We assumed ${v_{\rm ej} \approx 5 \; \rm m \; s^{-1}}$ for the collision on Figure \ref{fig: singleCollisionPositions} (compared to a collision speed ${v_{\rm col} \sim 100 \; \rm m \; s^{-1}}$), but estimates of $v_{\rm ej}$ are very uncertain. We chose $v_{\rm ej}$ based on arguments in \cite{Wyatt2002}. For a catastrophic collision between two equal-size bodies in the strength regime, 
 
\begin{equation}
v_{\rm ej} \approx \frac{1}{2} v_{\rm col} \left[\frac{f_{\rm KE}}{1 - \frac{1}{2} \left(\frac{2 Q_{\rm D}^*}{v_{\rm col}^2}\right)^{1.24}}\right]^{1/2}
\label{eq: ejectaVelFromCatastrophicCollision}
\end{equation}
 
\noindent where ${f_{\rm KE}}$ is the fraction of impact energy that ends up as kinetic energy of fragments, and ${Q_{\rm D}^*}$ quantifies the size-dependent material strength. The appropriate value of ${f_{\rm KE}}$ is uncertain, but we use a typical literature estimate of ${f_{\rm KE} \approx 0.1}$ (e.g. \citealt{Kenyon1999, Wyatt2002, Krivov2003}). ${Q_{\rm D}^*}$ is also uncertain, so we use a lower bound of ${1 \; \rm J \; kg^{-1}}$ (that of ${100 \; \rm m}$ diameter basalt; \citealt{Wyatt2002}). This yields a lower estimate of ${v_{\rm ej}}$ of ${20 \; \rm m \; s^{-1}}$. Given the uncertainties on the parameters, we adopt an even more conservative value of ${v_{\rm ej} \approx 5 \; \rm m \; s^{-1}}$, which is still large enough for dust to shear into a ring before other dynamical effects occur. Indeed, Equation \ref{eq: shearingTime} shows that provided ${v_{\rm ej} \gtrsim 0.7 \; \rm m \; s^{-1}}$, millimetre dust at ${142 \; \rm au}$ would shear out faster than the collision lifetime of ${10^6 \; \rm yr}$. It is therefore likely that ejecta released in the collision considered here would shear into a smooth ring before being collisionally depleted or disrupted by Fomalhaut b.

In this section we showed that, whilst the distribution of parent bodies in a resonant ring is expected to be broad and clumpy, the dust produced in a single collision between such bodies is expected to be narrow and azimuthally smooth. The specific dust morphology depends on the positions and velocities of the two parent bodies at the moment of collision; to predict the overall dust morphology arising from a debris ring in resonance with Fomalhaut b, we must therefore consider the superposition of dust released from multiple collisions.

\subsubsection{Overall dust morphology from multiple collisions}
\label{subsec: dustFromMultipleCollisions}

\noindent We now show that the overall dust morphology expected from the resonant model can well-resemble the observed Fomalhaut disc. We consider dust released in collisions between parent bodies from the simulation shown on Figure \ref{fig: fomcSimSkyPosAVsE}, just as we did for a single collision in Section \ref{subsec: dustFromSingleCollision}; the difference is that we now consider many collisions, and sum the resulting dust distributions to predict the overall dust morphology.

\begin{figure}
  \centering
   \includegraphics[width=7cm]{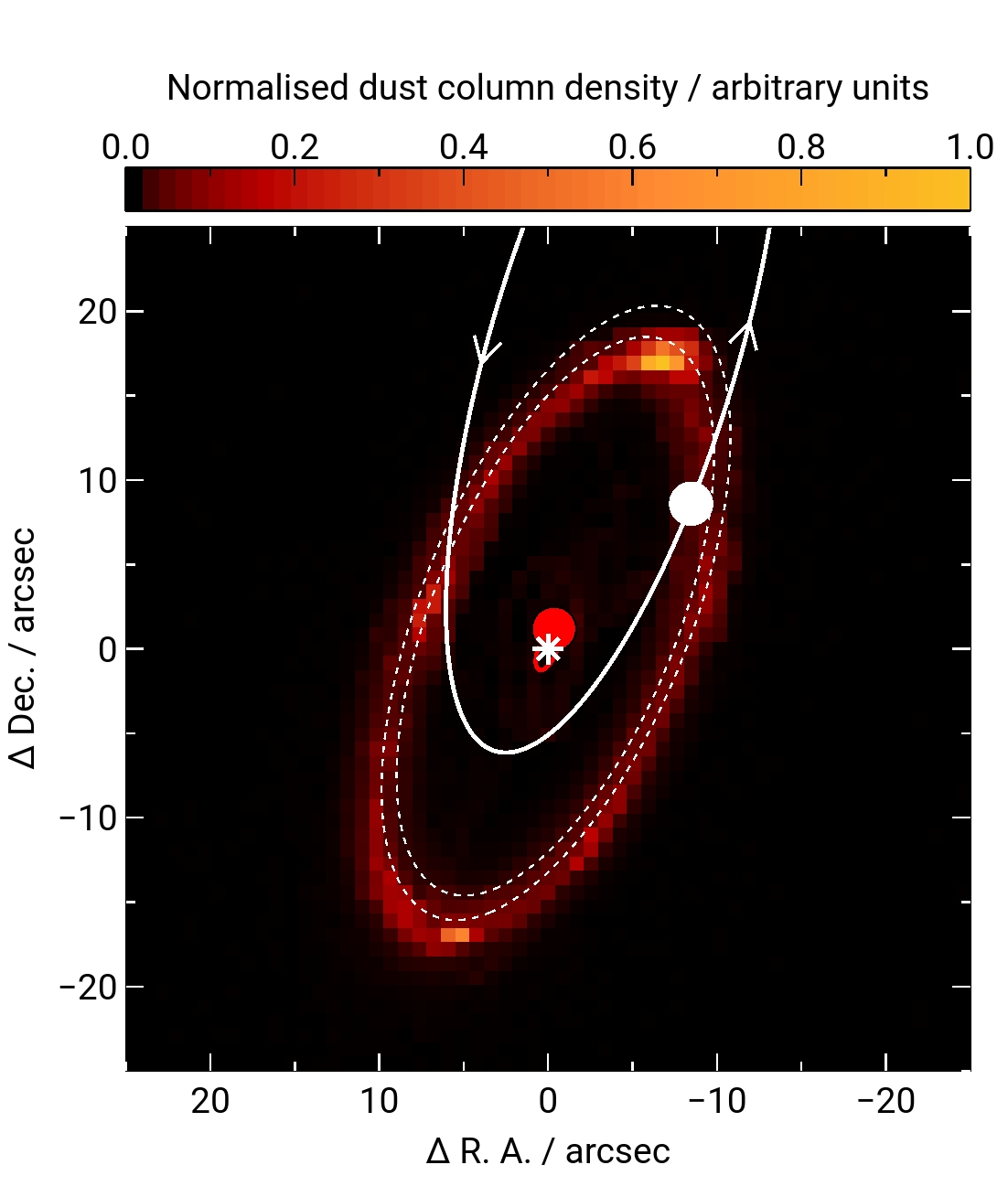}
   \caption{Simulated morphology of millimetre dust released in collisions between parent bodies, as described in Section \ref{subsec: dustFromMultipleCollisions}. The system comprises the star, debris, a ${0.1 \rm \; M_{Jup}}$ (${30 \; {\rm M_\oplus}}$) Fomalhaut b, and a hypothetical ${1 \rm \; M_{Jup}}$ planet at ${10 \; \rm au}$. The image shows the first observation epoch of Fomalhaut b (in 2004), when the simulated parent bodies have the distribution shown on Figures \ref{fig: fomcSimSkyPosAVsE} and \ref{fig: fomcSkyPosSepRes}. Shading denotes the column density of simulated dust particles released in collisions over the past ${1 \; \rm Myr}$ (the estimated lifetime of millimetre grains). Dashed white lines are the FWHM disc edges observed by ALMA \citep{Macgregor2017}. The figure shows that if Fomalhaut b sculpts debris through a resonant interaction, then the width, shape and orientation of the predicted millimetre dust distribution are similar to observations. Whilst the deprojected eccentricity of this particular dust ring (${0.05}$) is smaller than that of the observed disc (${0.12}$), the diversity of orbits allowed for Fomalhaut b (Figure \ref{fig: fomalhautbElementsInRegionOfInterest}) make it likely that there exist unexplored orbits or masses that better reproduce the observed debris morphology. The simulated dust distribution is also not as smooth as observations, owing to recent collisions where dust has not yet sheared out (a notable example is to the north); a slightly different collisional model could potentially resolve this discrepancy.}
   \label{fig: collisionsHeatmap}
\end{figure}

As shown in Section \ref{subsec: dustFromSingleCollision}, millimetre dust released in a single collision would quickly shear into a ring, before collisionally depleting over ${\sim 1 \; \rm Myr}$. To predict the present-day dust morphology, we take the parent body simulation shown on Figure \ref{fig: fomcSimSkyPosAVsE} and consider collisions that would have occurred within the last ${1 \; \rm Myr}$ (i.e. between simulation times of 439 and ${440 \; \rm Myr}$). During this time window we examine 200 simulation snapshots, each separated by ${5000 \; \rm yr}$. For each snapshot we search for pairs of parent bodies that lie within ${3 \; \rm au}$ of each other at the snapshot time, and treat each such close approach as a collision (${3 \; \rm au}$ is arbitrary, chosen to provide a reasonable number of close approaches). For each collision we then release 1000 dust particles into the simulation at the time of that collision, with the mean position and velocity of the two colliding bodies, plus a uniform ${5 \; \rm m \; s^{-1}}$ velocity dispersion as described in Section \ref{subsec: dustFromSingleCollision}. Each simulation is then advanced from the collision time until now (i.e. up to ${440 \; \rm Myr}$), including the perturbing effects of Fomalhaut b and the hypothetical inner planet. Finally, we take the dust distributions from all of these individual collisions (${\sim 100}$ collisions in total) at the present day, and sum them to produce the predicted overall dust morphology. We show this morphology on Figure \ref{fig: collisionsHeatmap}.

The figure shows that the width, shape and orientation of millimetre dust predicted from the resonant model can be similar to the observed disc around Fomalhaut. In particular, note that the dust distribution is substantially smoother than the parent body distribution on Figures \ref{fig: fomcSimSkyPosAVsE} and \ref{fig: fomcSkyPosSepRes}, and so dust better resembles the observed disc than the resonant parent bodies do. As described in Section \ref{subsec: dustFromSingleCollision}, the relative smoothness of dust arises from grains shearing out long before they are perturbed by Fomalhaut b or the hypothetical inner planet(s). The dust distribution is also more radially peaked than that of the parent bodies; since collisions between parent bodies mainly occur in the centres of resonant clumps, most dust grains occupy orbits tracing the centres of these clumps. This is in agreement with previous works, that suggest collision rates in the centres of resonant clumps are potentially an order of magnitude higher than those outside \citep{Wyatt2006, Queck2007}. 

Whilst the dust distribution on Figure \ref{fig: collisionsHeatmap} well-resembles the Fomalhaut disc, there are also a few differences between this specific simulation and observations. Firstly, the deprojected eccentricity of this particular dust ring (${0.05}$) is smaller than that of the observed disc (${0.12}$). This is not a major problem, because the simulated disc eccentricity is set by the chosen parameters of Fomalhaut b. The high-eccentricity resonant interaction is complex, and small parameter changes can cause very different particle behaviours (Figure \ref{fig: 5_3And7_4PhaseMaps}). This, combined with the diversity of possible orbits of Fomalhaut b (Figure \ref{fig: fomalhautbElementsInRegionOfInterest}), make it likely that some alternative parameters could better reproduce observed debris. Conversely, not all resonant interactions result in apsidally aligned discs; for some simulations, misaligned lobes like those on Figure \ref{fig: removalSimsPosAvsE} become more pronounced, and the disc may have a more complicated morphology. A detailed parameter space search is therefore required to identify setups that best reproduce observations, which could constrain the orbit and mass of Fomalhaut b if it sculpts the disc through resonant interactions. However, the high computational cost of such a search takes it beyond the scope of this paper.

A second difference is that our simulated dust morphology is not as smooth as the ALMA image. This can be seen from the various overdensities in the simulated dust (Figure \ref{fig: collisionsHeatmap}), where some regions have up to three times the mean column density of the ring. These local dust overdensities were produced in recent collisions, and have not yet sheared out. However, the quantity and magnitude of these overdensities are determined by the specific parameters employed in our collisional model, and the use of different parameters would change the smoothness of the dust ring. In particular, changing the arbitrary close approach distance of ${3 \; \rm au}$ would affect the number of simulated collisions and the initial orbits of the released dust. Similarly, a different ejecta velocity dispersion ${\Delta v}$ would affect the dust shearing timescale, and hence the prominence of dust overdensities from recent collisions. A more detailed exploration of these collisional parameters is beyond the scope of this paper, but it is likely that there is tolerance within the collisional model to generate a smoother dust ring than that on Figure \ref{fig: collisionsHeatmap}. Also, aside from very recent collisions, the simulated dust ring has relatively uniform density; this differs from the `apocentre glow' observed in reality, where millimetre grains moving more slowly at apocentre cause the north-west side of the disc to be brighter than the south-east side \citep{Pan2016, Macgregor2017}. This discrepancy is caused by our example simulated ring being less eccentric than the observed disc, but a different simulation with a more eccentric dust morphology should reproduce the observed apocentre glow.

In summary, our resonant model predicts that large parent bodies in the Fomalhaut disc would have a broad, clumpy distribution (Figure \ref{fig: fomcSimSkyPosAVsE}), whilst dust may have a narrower, smoother morphology that better reproduces the observed disc (Figure \ref{fig: collisionsHeatmap}). A potential observational test of this scenario would therefore be to compare the dust distribution to that of parent bodies. Whilst such large bodies cannot be directly detected at present, it may be possible to infer information about them through observations of background stars transiting behind the disc; such an opportunity may arise in the near future \citep{Zeegers2014, Meshkat2016}. These observations would also be useful for constraining the total mass of the Fomalhaut ring, which would have further, general implications for debris disc evolution models.

\section{Discussion}
\label{sec: discussion}

\noindent We have shown that even if Fomalhaut b is massive and coplanar with the debris disc, then resonant particles on orbits similar to the observed dust ring can be stable for the system lifetime. This argues against the idea that Fomalhaut b cannot have significant mass without disrupting the observed debris disc. We also showed that the expected millimetre dust morphology arising from collisions between resonant bodies can well-match the width, shape and orientation of the observed dust ring. We now discuss the Fomalhaut system parameters required for our scenario to operate, the effects of non-negligible disc mass, the nature of Fomalhaut b, and potential implications of the scenario for the excess near-infrared emission observed in the system.

\subsection{System parameters required to reproduce the observed disc}
\label{subsec: requiredSystemParameters}

We explored a scenario where a massive, coplanar Fomalhaut b is placed onto an observationally allowed orbit that crosses an unexcited debris disc, resulting in the removal of non-resonant particles and the survival of long-term-stable bodies on orbits similar to the observed disc. We now discuss the system parameters required for this scenario to occur.

Firstly, let us consider the required mass of Fomalhaut b. If Fomalhaut b were too massive then it would scatter and eject almost all resonant debris, and our simulations show that Fomalhaut b would probably have to be less massive than ${1 \;\rm M_{Jup}}$ (${300 \; \rm M_\oplus}$) for our scenario to occur (Table \ref{tab: simulationsRun}). Conversely, it must be massive enough to drive non-resonant debris into the inner system for removal by inner planets within the stellar lifetime; this requires the secular time to be less than twice the stellar lifetime (since particle eccentricities are maximised after half a secular time), and so Fomalhaut b must be at least ${1 \; \rm M_\oplus}$ for this scenario to take place (Equation \ref{eq: secularTimescale}). Of course, such a low-mass Fomalhaut b would almost certainly be perturbed by the disc (Section \ref{subsec: discMass}), so its mass should also be greater than the mass of the observed disc if it is to dominate debris evolution.

Secondly, we must consider the required orbital plane of Fomalhaut b. We assumed that Fomalhaut b orbits close to the disc midplane, as expected from system formation and evolution arguments. This is consistent with the sky-plane motion of Fomalhaut b; if it moves under gravity alone, then a significant fraction of observationally allowed orbits lie within $20^\circ$ of the disc plane (Figure \ref{fig: fomalhautbAllowedOrbits}). However, higher inclinations are also allowed. We have not explored these highly inclined orbits in detail, but we did run 10 dynamical simulations with Fomalhaut b on observationally allowed orbits with inclinations between ${30}$ and ${130^\circ}$. None of these simulations reproduced the observed disc. Debris in these interactions has more vertical structure than in the coplanar case, which does not match observations (the observed ring has a small opening angle of ${1.00 \pm 0.25^\circ }$; \citealt{Boley2012}). Also, far fewer bodies are resonant in the inclined simulations, potentially because resonant trapping in such extreme configurations is less efficient than in the coplanar case (e.g. \citealt{Pearce2014}). Whilst it is possible that an unexplored region of parameter space could allow Fomalhaut b to be highly inclined to a resonant disc that matches the observed ring, we suspect that Fomalhaut b must orbit close to the disc plane if the scenario we explore is to work.

Thirdly, we require at least one planet interior to Fomalhaut b to remove much of the high-eccentricity debris. This is because the observed disc is unlikely to be resonant unless the mass of Fomalhaut b is less than ${1 \;\rm M_{Jup}}$, and in this case Fomalhaut b alone cannot clear all unstable debris within the system lifetime. An additional planet is not an unreasonable requirement; regardless of whether the debris ring is resonant, it is likely that Fomalhaut b is not the only body in the system for several reasons. First, many stellar systems (including our own) are known to host more than one planet, so multi-planet systems may not be unusual. Second, broad cavities between debris populations (such as the ${\sim 100 \; \rm au}$ gap between the imaged debris ring and the inferred warm belt in the Fomalhaut system) are most-simply explained by multi-planet clearing \citep{Faber2007, Su2013}. Finally, if Fomalhaut b is massive then a large body is probably required to drive it onto its eccentric orbit \citep{Faramaz2015}. These arguments suggest that additional large bodies are expected in the system, as required by our model. However, we again stress that the hypothetical inner planet examined in Section \ref{subsec: removingUnstableDebris} is just one example of a planet capable of removing unstable debris in our model; a planet with a different mass at a different location, or multiple planets, would have the same effect. Our only constraint is that we require at least one body to reside in the inner system to remove high-eccentricity debris.

Finally, we require an initial planetesimal disc that can populate resonances with debris. This is potentially a problem with our scenario; given the low efficiency of resonance trapping (less than ${1 \; \rm per \; cent}$ of bodies initially present in our simulation on Figure \ref{fig: fomcSimSkyPosAVsE} are in resonance after ${440 \; \rm Myr}$), the original disc may have had to be very massive to yield sufficient resonant material. For example, if less than ${1 \; \rm per \; cent}$ of the original debris disc remains, and the lowest estimates of the current debris mass are ${2-30 \; {\rm M_\oplus}}$ \citep{Wyatt2002, Chiang2009, Boley2012, Krivov2021}, then the original disc would have been at least of order ${100-1000 \; {\rm M_\oplus}}$; this would alter our modelled interaction dynamics and may even be unphysically large (see Section \ref{subsec: discMass}). However, there are several possibilities that could reduce the requirement for such a high initial disc mass, or lessen its effect on Fomalhaut b. Firstly, if Fomalhaut b did not rapidly transition to its current orbit but instead migrated gradually, then it could have swept additional material into resonance and therefore preserve more than ${1 \; \rm per \; cent}$ of the original disc mass (e.g. \citealt{Faramaz2014}). Another possibility is that Fomalhaut b transitioned to its current orbit more recently than ${440 \; \rm Myr}$ ago, in which case a greater fraction of resonant debris would be preserved (some of this resonant debris would be ejected in the future). Alternatively, if Fomalhaut b transitioned to its current orbit during a major dynamical upheaval of the system, then several planetary bodies may have undergone significant migration simultaneously; multiple bodies may have quickly cleared debris, reducing the influence of an initially massive disc on the orbit of Fomalhaut b. Whilst these and other scenarios are possible, we admit that the potential requirement for a high initial disc mass is a concern for the feasibility of our resonant model. Our main conclusion still holds (the observed debris ring can be stable if in resonance with a massive Fomalhaut b, regardless of the origin of that debris), but further exploration of possible evolution pathways is warranted to determine whether the need for a massive initial disc can be negated.

\subsection{Dynamical effect of disc mass}
\label{subsec: discMass}

Different assumptions about the mass of the Fomalhaut debris disc could affect the dynamical interaction studied in this paper. Our simulations assumed a massless disc, which is expected to produce reliable results provided that the perturbing planet is more massive than the disc. If this were not the case, then the planet could undergo significant orbital evolution that would fundamentally change the interaction (e.g. \citealt{Pearce2014, Pearce2015PltDsc}). Since we find that resonance trapping works best if Fomalhaut b has a mass of ${\sim 0.1 \; {\rm M_{Jup}}}$, the Fomalhaut disc may have to be less than ${\sim 0.1 \; {\rm M_{Jup}}}$ (a few tens of Earth masses) for the interaction studied here to occur. However, the only thing we can be relatively sure of is that the dust mass is of order ${10^{-2} \; {\rm M}_\oplus}$ \citep{Macgregor2017}; the total disc mass must be calculated from this, and is therefore uncertain. Our requirement for a total disc mass of at most a few tens of Earth masses is compatible with the lower estimates of ${2-30 \; {\rm M_\oplus}}$ for the current Fomalhaut debris ring \citep{Wyatt2002, Chiang2009, Boley2012, Krivov2021}, but is much smaller than other estimates of ${100 \; {\rm M}_\oplus}$ or more (e.g. \citealt{Acke2012}). However, \citet{Krivov2021} argue that such high masses may be unreasonable for debris discs, and that disc masses could be systematically overestimated. Given the range of mass estimates, it is possible that the Fomalhaut disc mass is small enough for the interaction studied here to occur.

Alternatively, the self gravity of a massive disc could be sufficient for the disc to resist deformation by a planet (\citealt{Goldreich1979, Sefilian2020}; L\"{o}hne et al. in prep). Since we find that a ${1 \; {\rm M_{Jup}}}$ Fomalhaut b would rapidly remove a massless resonant disc, the inclusion of self gravity could mean that debris better resists ejection, in which case the resonant interaction studied here could work even if Fomalhaut b has a mass of ${1 \; {\rm M_{Jup}}}$. In turn, this would allow the debris disc to be more massive than ${0.1 \; {\rm M_{Jup}}}$ (${30 \; {\rm M_\oplus}}$) without significantly perturbing Fomalhaut b. However, the very high computational cost of including self gravity in a broad disc where resonant, secular and scattering interactions are all important puts it beyond the scope of this paper. In light of the uncertainties on disc mass and self gravity we chose to only model massless discs, but we stress that the interaction could be different from that presented here if disc mass is taken into account.

\subsection{The nature of Fomalhaut b}
\label{subsec: natureOfFomb}

We have shown that if Fomalhaut b is less massive than Jupiter then it could sculpt the observed debris ring. This means that very few observationally allowed masses for Fomalhaut b can be dynamically excluded. A distinction between high- and low-mass models is therefore difficult through dynamical arguments alone, although one possibility is that a massive Fomalhaut b passing through the ring might leave some signature that could potentially be observed.

The orbit of Fomalhaut b is also difficult to narrow down, as a wide range of bound and unbound trajectories are possible. If Fomalhaut b moves under gravity alone, then the best orbital constraints would come from the observation of orbital curvature, which would exclude a large number of orbits from Figure \ref{fig: fomalhautbAllowedOrbits}. It is also possible to employ the following probabilistic argument in favour of low semimajor axes; whilst this does not exclude high semimajor axes or unbound solutions, it can be used to argue that they are less likely. If Fomalhaut b were massive and on the observationally allowed orbit used as an example in this paper (coplanar with the disc, with semimajor axis ${203 \; \rm au}$ and eccentricity 0.760), then it will cross the centre of the ring in projection when its true anomaly reaches ${124^\circ}$. This would occur in the year 2047. It would later come back inside the ring in projection at true anomaly ${232^\circ}$ which means that, given an orbital period of ${2000 \; \rm yr}$, Fomalhaut b would spend ${17 \; \rm per \; cent}$ of its time inside the ring in projection. It is therefore not unreasonable to believe that Fomalhaut b occupies this or a similar orbit, and we happen to have observed the system while Fomalhaut b lies interior to the ring. Conversely, if Fomalhaut b has a much larger semimajor axis, or is unbound and escaping, then the probability of observing it inside the ring rapidly diminishes. This argument supports Fomalhaut b having as small a semimajor axis as observationally allowed, and our simulations generally show that the smaller the semimajor axis, the more debris is able to survive on low-eccentricity, perturber-crossing orbits (Table \ref{tab: simulationsRun}).

If Fomalhaut b is of planetary mass, then it would need something like a ring or circumplanetary dust cloud to increase its brightness in scattered light \citep{Kalas2008, Kennedy2011, Currie2012, Galicher2013, Kenyon2014, Tamayo2014}. An alternative hypothesis is that Fomalhaut b does not have significant mass, but is in fact a dispersing dust cloud \citep{Janson2012, Galicher2013, Lawler2015, Janson2020}; in this case, the disc eccentricity would have to be driven by a different mechanism (e.g. \citealt{Shannon2014, Faramaz2015, Kaib2018, Kennedy2020}). We have shown that Fomalhaut b is not precluded from having planetary mass by dynamical arguments, so a distinction between the planet and dust cloud hypotheses must come from observations. We may be close to this; \cite{Gaspar2020} argue that Fomalhaut b may be fading and experiencing non-gravitational acceleration, which they argue favours an escaping dust cloud model, although additional observations are required to confirm these results. 

The probability of observing an escaping dust cloud would be low, given how rapidly it would leave the system (unless such clouds were frequently created). However, this argument does not rule out the escaping dust cloud hypothesis. Instead we should consider whether fading and non-gravitational acceleration could distinguish between planetary and dust cloud models. Fading alone would not necessarily preclude Fomalhaut b from having significant mass, because it is only observed in visible light, and one could imagine various reasons for its visible flux to vary. If Fomalhaut b is on our example orbit with semimajor axis ${203 \; \rm au}$ and eccentricity 0.760, and presents a constant area to the star, then it would have faded in scattered light by ${10 \; \rm per \; cent}$ between 2004 and 2014 as it moves outwards. More dramatic fading could also be consistent with a circumplanetary dust model; for example, a recent collision within a circumplanetary dust cloud, a tilted or precessing ring system, or the obscuration or evolution of circumplanetary dust clumps would all cause brightness variations, and so fading alone is not sufficient evidence that Fomalhaut b has negligible mass. However, non-gravitational acceleration would be much more difficult to reconcile with a massive companion, and if it were definitively shown that Fomalhaut b cannot be moving under gravity alone then this would effectively rule out a planetary explanation. The best way to do this is to take more observations of Fomalhaut b over a long period, and demonstrate that its observed orbital curvature (or lack of it) is incompatible with any trajectories set by gravity alone. We therefore urge more observations to be made of Fomalhaut b, in order to search for non-gravitational acceleration that would distinguish between the planetary and dust cloud hypotheses.

\subsection{Hot dust in the Fomalhaut system}
\label{subsec: hotDust}

An interesting final point is that, since we have shown that a massive Fomalhaut b on an observationally allowed orbit would drive non-resonant debris deep into the inner system, our model could have implications for the excess near-infrared emission detected around the star \citep{Absil2009, Mennesson2013, Lebreton2013}. Such emission is detected for a significant fraction of  main-sequence stars, and is typically ascribed to sub-micron grains at very small stellocentric distances (e.g. \citealt{Absil2006, Kral2017, Kirchschlager2017}). However, the nature of this material is mysterious, since such grains are expected to sublimate or blow away too rapidly to reproduce observations \citep{vanLieshout2014, Sezestre2019}. The two most-promising solutions are either that grains are trapped close to such stars \citep{Rieke2016, Kimura2020, Pearce2020}, or that material is continuously resupplied by star-grazing debris \citep{Bonsor2013, Bonsor2014, Raymond2014, Faramaz2017, Sezestre2019}. Both models require material to get very close to the star from an external source, and we have shown that a massive Fomalhaut b on an observationally allowed orbit is capable of driving debris down to the required distances (${\sim 0.1 \; \rm au}$). Also, since this is a secular process, such an inflow could be sustained for a very long time. Whilst a detailed examination of hot dust supply is beyond the scope of this paper, it is interesting that a consequence of our model (that some debris from the outer regions should reach the sublimation region of Fomalhaut) is consistent with the presence of a near-infrared excess in the system.
  

\section{Conclusions}
\label{sec: conclusions}

\noindent We re-examine the orbit of Fomalhaut b and show that, if it moves under gravity alone and orbits within $5^\circ$ of the debris disc plane, then it is likely bound to the star and apsidally aligned with the disc to within $20^\circ$. Its pericentre and apocentre would be interior and exterior to the disc respectively, and it would have an eccentricity of at least $0.76$. Such an extreme orbital configuration could be expected to disrupt the observed debris ring, and this is often used as an argument that Fomalhaut b cannot have significant mass. However, we model the interaction between debris and a massive Fomalhaut b on these observationally allowed orbits, and show that debris populations can be stable for the stellar lifetime if in internal mean-motion resonances with an intermediate-mass, coplanar Fomalhaut b. This debris can have eccentricities and semimajor axes that are similar to the observed debris ring. Whilst these resonant parent bodies would have a clumpy distribution, we show that millimetre dust created in collisions between these bodies would have a much smoother morphology that could reproduce the observed disc. In particular, the width, shape and orientation of our simulated dust distribution compare well to observations. Our resonant scenario could operate provided that Fomalhaut b is less massive than Jupiter, so the argument that Fomalhaut b cannot have significant mass without disrupting the observed disc is not necessarily correct. We also show that Fomalhaut b is expected to drive debris deep into the inner system, which is potentially consistent with detections of excess near-infrared emission ascribed to material very close to the star.

More generally, the stable, narrow, low-eccentricity debris rings that we simulate may not have been expected to exist in the presence of a highly eccentric perturber, but we find that they manifest themselves across a broad range of parameter space. Our results could therefore have other implications beyond the Fomalhaut system, as a new mechanism by which a planet could sculpt a narrow, eccentric debris disc.


\section*{Acknowledgements}
\noindent We thank the anonymous referee for their helpful comments. TDP, MB, AVK, TL and PPP are supported by the \textit{Deutsche Forschungsgemeinschaft} (DFG), grants ${\rm Kr \; 2164/13-2}$, ${\rm Kr \; 2164/14-2}$, ${\rm Kr \; 2164/15-2}$ and ${\rm Lo \; 1715/2-2}$. VF's postdoctoral fellowship is supported by the Exoplanet Science Initiative at the Jet Propulsion Laboratory, California Institute of Technology, under a contract with the National Aeronautics and Space Administration (80NM0018D0004).

\section*{Data availability}
\noindent The data underlying this article will be shared upon reasonable request to the corresponding author.


\bibliographystyle{mn2e}
\bibliography{bib_fomb}


\appendix

\section{Table of simulations}
\label{app: simulationsRun}

Table \ref{tab: simulationsRun} shows the parameters of Fomalhaut b examined in our simulations. We only list simulations that contain the star, large debris and Fomalhaut b, and of these, only simulations where the latter orbits in the disc midplane (additional simulations with an inclined Fomalhaut b, or those including dust or a hypothetical inner planet, are not listed).

\begin{table*}
\begin{tabular}{c c c c c c}
\hline
Sim. Number & ${M_{\rm plt} / \rm M_{Jup}}$ & ${a_{\rm plt}} / \rm au$ & ${e_{\rm plt}}$ & ${f_{\rm stable}}$ & ${f_{\rm unexcited}}$ \\
\hline
1	&	0.01	&	170	&	0.80	&	0.092	&	0.011	\\
2	&	\texttt{"}	&	187	&	0.76	&	0.121	&	0.020	\\
3	&	\texttt{"}	&	\texttt{"}	&	0.84	&	0.105	&	0.011	\\
4	&	\texttt{"}	&	\texttt{"}	&	0.80	&	0.123	&	0.027	\\
5	&	\texttt{"}	&	203	&	0.76	&	0.121	&	0.013	\\
6	&	\texttt{"}	&	\texttt{"}	&	0.85	&	0.114	&	0.015	\\
7	&	\texttt{"}	&	\texttt{"}	&	0.81	&	0.116	&	0.016	\\
8	&	\texttt{"}	&	226	&	0.76	&	0.205	&	0.019	\\
9	&	\texttt{"}	&	\texttt{"}	&	0.87	&	0.095	&	0.011	\\
10	&	\texttt{"}	&	\texttt{"}	&	0.82	&	0.133	&	0.012	\\
11$^\dagger$	&	\texttt{"}	&	296	&	0.80	&	0.358	&	0.026	\\
12$^\dagger$	&	\texttt{"}	&	\texttt{"}	&	0.89	&	0.079	&	0.021	\\
13$^\dagger$	&	\texttt{"}	&	\texttt{"}	&	0.85	&	0.141	&	0.020	\\
14$^\dagger$	&	\texttt{"}	&	400	&	0.88	&	0.210	&	0.018	\\
15$^\dagger$	&	\texttt{"}	&	500	&	0.90	&	0.349	&	0.022	\\
16$^\dagger$	&	\texttt{"}	&	600	&	0.92	&	0.406	&	0.030	\\
17$^\dagger$	&	\texttt{"}	&	700	&	0.93	&	0.430	&	0.045	\\
18$^\dagger$	&	\texttt{"}	&	800	&	0.94	&	0.463	&	0.064	\\
19$^\dagger$	&	\texttt{"}	&	900	&	0.95	&	0.506	&	0.152	\\
20$^\dagger$	&	\texttt{"}	&	1000	&	\texttt{"}	&	0.549	&	0.380	\\
\hline											
21	&	0.1	&	170	&	0.80	&	0.017	&	0.017	\\
22	&	\texttt{"}	&	187	&	0.76	&	0.014	&	0.011	\\
23	&	\texttt{"}	&	\texttt{"}	&	0.84	&	0.019	&	0.019	\\
24	&	\texttt{"}	&	\texttt{"}	&	0.80	&	0.024	&	0.022	\\
25	&	\texttt{"}	&	203	&	0.76	&	0.008	&	0.003	\\
26	&	\texttt{"}	&	\texttt{"}	&	0.85	&	\texttt{"}	&	0.008	\\
27	&	\texttt{"}	&	\texttt{"}	&	0.81	&	0.009	&	0.009	\\
28	&	\texttt{"}	&	226	&	0.76	&	0.001	&	0.001	\\
29	&	\texttt{"}	&	\texttt{"}	&	0.87	&	0.006	&	0.005	\\
30	&	\texttt{"}	&	\texttt{"}	&	0.82	&	0.000	&	0.000	\\
31	&	\texttt{"}	&	296	&	0.80	&	0.007	&	0.006	\\
32	&	\texttt{"}	&	\texttt{"}	&	0.89	&	0.004	&	0.004	\\
33	&	\texttt{"}	&	\texttt{"}	&	0.85	&	\texttt{"}	&	0.001	\\
34	&	\texttt{"}	&	400	&	0.88	&	0.008	&	0.004	\\
35$^\dagger$	&	\texttt{"}	&	500	&	0.90	&	0.001	&	0.001	\\
36$^\dagger$	&	\texttt{"}	&	600	&	0.92	&	0.002	&	0.000	\\
37$^\dagger$	&	\texttt{"}	&	700	&	0.93	&	0.004	&	0.003	\\
38$^\dagger$	&	\texttt{"}	&	800	&	0.94	&	0.006	&	0.001	\\
39$^\dagger$	&	\texttt{"}	&	900	&	0.95	&	0.002	&	\texttt{"}	\\
40$^\dagger$	&	\texttt{"}	&	1000	&	\texttt{"}	&	0.001	&	\texttt{"}	\\
\hline											
41	&	1	&	170	&	0.80	&	0.000	&	0.000	\\
42	&	\texttt{"}	&	187	&	0.76	&	\texttt{"}	&	\texttt{"}	\\
43	&	\texttt{"}	&	\texttt{"}	&	0.84	&	0.001	&	0.001	\\
44	&	\texttt{"}	&	\texttt{"}	&	0.80	&	0.000	&	0.000	\\
45	&	\texttt{"}	&	203	&	0.76	&	\texttt{"}	&	\texttt{"}	\\
46	&	\texttt{"}	&	\texttt{"}	&	0.85	&	\texttt{"}	&	\texttt{"}	\\
47	&	\texttt{"}	&	\texttt{"}	&	0.81	&	\texttt{"}	&	\texttt{"}	\\
48	&	\texttt{"}	&	226	&	0.76	&	\texttt{"}	&	\texttt{"}	\\
49	&	\texttt{"}	&	\texttt{"}	&	0.87	&	\texttt{"}	&	\texttt{"}	\\
50	&	\texttt{"}	&	\texttt{"}	&	0.82	&	\texttt{"}	&	\texttt{"}	\\
51	&	\texttt{"}	&	296	&	0.80	&	0.003	&	0.003	\\
52	&	\texttt{"}	&	\texttt{"}	&	0.89	&	0.000	&	0.000	\\
53	&	\texttt{"}	&	\texttt{"}	&	0.85	&	\texttt{"}	&	\texttt{"}	\\
54	&	\texttt{"}	&	400	&	0.88	&	\texttt{"}	&	\texttt{"}	\\
55	&	\texttt{"}	&	500	&	0.90	&	\texttt{"}	&	\texttt{"}	\\
56	&	\texttt{"}	&	600	&	0.92	&	\texttt{"}	&	\texttt{"}	\\
57	&	\texttt{"}	&	700	&	0.93	&	\texttt{"}	&	\texttt{"}	\\
58	&	\texttt{"}	&	800	&	0.94	&	\texttt{"}	&	\texttt{"}	\\
59	&	\texttt{"}	&	900	&	0.95	&	\texttt{"}	&	\texttt{"}	\\
60$^\dagger$	&	\texttt{"}	&	1000	&	\texttt{"}	&	\texttt{"}	&	\texttt{"}	\\

\hline
\end{tabular}
\caption{All simulations in this paper that contain only the star, large debris and Fomalhaut b, where the latter initially orbits in the disc midplane. The value ${f_{\rm stable}}$ is the fraction of initial particles whose semimajor axes change by less than ${2 \rm \; per \; cent}$ over the ${440 \; \rm Myr}$ simulation. The value ${f_{\rm unexcited}}$ is the fraction of particles whose semimajor axes change by less than ${2 \rm \; per \; cent}$ and their eccentricity and inclination never exceed 0.4 and ${20^\circ}$, respectively. Entries marked \texttt{"} are identical to the above value in that column. In simulations marked $^\dagger$ the secular timescale is over twice the system age, so our scenario is disfavoured for those setups.}
\label{tab: simulationsRun}
\end{table*}

\section{Derivation of collisional ejecta shearing timescale}
\label{app: shearingTimeDerivation}
\noindent Suppose that a large body is on a circular orbit with radius $r$ and velocity $v_{\rm K}$, and it releases a cloud of particles with an additional velocity dispersion ${\Delta v}$. Also suppose that these particles are large enough to be unaffected by radiation pressure, and that ${\Delta v \ll v_{\rm K}}$. The slowest and fastest particles will therefore have speeds ${v_1 \equiv v_{\rm K} - \Delta v}$ and ${v_2 \equiv v_{\rm K} + \Delta v}$ respectively. The particles will all have different semimajor axes $a$, where the semimajor axis of the slowest particle is

\begin{equation}
a_1 = \left( \frac{2}{r} - \frac{v_1^2}{G M_*} \right)^{-1}
\label{eq: semimajorAxis}
\end{equation}

\noindent (with a similar expression for $a_2$, the semimajor axis of the fastest particle). The particles will therefore progress around their orbits at different rates. After several orbits their positions will become uncorrelated, and they will shear out into a smoother distribution.

As each particle progresses around its orbit, its mean anomaly $\mathcal{M}$ increases with the time since release $t$ as ${\mathcal{M} = n t + \mathcal{M}_0}$, where ${n = \sqrt{G M_* / a^3}}$ is its mean motion and $\mathcal{M}_0$ its mean anomaly at time ${t=0}$. The difference between the mean anomalies of the fastest and slowest particles is therefore

\begin{equation}
\Delta \mathcal{M} = (n_1 - n_2) t + \Delta \mathcal{M}_0,
\label{eq: differenceInMeanAnomaliesOverTime}
\end{equation}

\noindent where ${n_1}$ and ${n_2}$ are the mean motions of the slowest and fastest particles respectively, and ${\Delta \mathcal{M}_0}$ is the difference between their initial mean anomalies (note that since ${v_2 > v_1}$, ${a_2 > a_1}$ and therefore ${n_2 < n_1}$).

Over time the positions of the fastest and slowest particles become less correlated, and once ${(n_1 - n_2) t \sim 2 \upi}$ the cloud of particles will have significantly sheared out. We therefore define the shearing time ${t_{\rm shear}}$ as the time when ${(n_1 - n_2) t_{\rm shear} \equiv 2 \upi}$. Substituting $n_1$ and $n_2$ from above, taking $a_1$ and $a_2$ from Equation \ref{eq: semimajorAxis}, and assuming ${\Delta v \ll v_{\rm K}}$, we arrive at the shearing time

\begin{equation}
t_{\rm shear} \sim \frac{r}{\Delta v}.
\label{eq: shearingTimeFromDerivation}
\end{equation}


\label{lastpage}

\end{document}